\documentclass[sn-mathphys-num]{sn-jnl}

\usepackage{graphicx}%
\usepackage{multirow}%
\usepackage{amsmath,amssymb,amsfonts}%
\usepackage{amsthm}%
\usepackage{mathrsfs}%
\usepackage[title]{appendix}%
\usepackage{xcolor}%
\usepackage{textcomp}%
\usepackage{manyfoot}%
\usepackage{booktabs}%
\usepackage{algorithm}%
\usepackage{algorithmicx}%
\usepackage{algpseudocode}%
\usepackage{listings}%
\usepackage{booktabs}

\theoremstyle{thmstyleone}%
\theoremstyle{thmstyletwo}%

\theoremstyle{thmstylethree}%

\begin{document}

\title[A Survey on the Use by Students of LLMs in Education Settings]{Is ChatGPT Massively Used by Students Nowadays? A Survey on the Use of Large Language Models such as ChatGPT in Educational Settings}


\author*[1]{\fnm{Jérémie} \sur{Sublime}}\email{jsublime@isep.fr}
\equalcont{These authors contributed equally to this work.}

\author[2]{\fnm{Ilaria} \sur{Renna}}\email{ilaria.renna@unitn.it}
\equalcont{These authors contributed equally to this work.}

\affil[1]{\orgname{Isep -- School of Digital Engineers}, \orgaddress{\street{10 rue de Vanves}, \postcode{92130}, \city{Issy-Les-Moulineaux}, \country{France}}}

\affil[2]{\orgdiv{Department of Economics and Management}, \orgname{University of Trento}, \orgaddress{\street{Via Inama 5}, \city{Trento}, \postcode{38122}, \country{Italy}}}


\abstract{The rapid adoption of Generative AI (GenAI) based on Large Language Models (LLMs) such as ChatGPT has recently and profoundly impacted education, offering transformative opportunities while raising significant concerns. In this study we present the results of a survey that investigates how 395 students aged 13 to 25 years old in France and Italy integrate LLMs into their educational routines.

Key findings include the widespread use of these tools across all age groups and disciplines, with older students and male students demonstrating higher usage frequencies, particularly in scientific contexts. The results also show gender disparities, raising concerns about an emerging AI literacy and technological gender gap. Additionally, while most students utilise LLMs constructively, the lack of systematic proofreading and critical evaluation among younger users suggests potential risks to cognitive skills development, including critical thinking and foundational knowledge. The survey results underscore the need for educational institutions to adapt their curricula to integrate AI tools effectively, promoting ethical use, critical thinking, and awareness of AI limitations and environmental costs. This paper provides actionable recommendations for fostering equitable and effective cohabitation of LLMs and education while addressing emerging challenges.}




\maketitle

\section{Introduction}

Few inventions and innovations have genuinely transformed education at large, particularly by enhancing access to knowledge. Notable among these are the advent of writing around 3300 BCE, which facilitated the transmission of knowledge across generations and cultures; the Gutenberg printing press in approximately 1440 CE, which greatly simplified the duplication and dissemination of ideas and knowledge, thereby encouraging wider literacy and education; the large-scale deployment of the World Wide Web in the late 1990s and early 2000s, which allowed for rapid, affordable, and accessible information sharing via the Internet, especially through online encyclopedias such as Wikipedia; and, more recently, the public emergence of Large Language Models (LLMs) \cite{minaee2024} in 2022, such as ChatGPT (Chat Generative Pre-Trained Transformer) \cite{openai2023chatgpt}, which have made information access even more straightforward.

However, LLMs differ from previous inventions that facilitated the spread of information and knowledge in several key ways \cite{burton2024large,luo2024large}. While \textit{writing}, the \textit{printing press}, and the \textit{Internet} primarily made information more accessible, LLMs provide an array of additional functions, such as multi-language translation, summarisation, simplification of complex information, and advanced writing capabilities to structure and organise content. In other words, LLMs assist people not only with accessing information but also with tasks traditionally considered cognitive. Consequently, these models may also be likened to inventions that support cognitive processes, such as \textit{calculators} and \textit{computers}, which are far more efficient than humans in computation and have significantly reduced the need for mental arithmetic resulting, as a side effect, in a decline in our capacity for mental mathematics.

Whether we welcome them or not, LLMs have arrived and they are here to stay. It is however important to understand how they can be exploited in a positive way while minimizing negative impacts.
Their impact on industry, particularly within the research and innovation sectors, is already evident and largely positive. When humans and AI collaborate, the gains in productivity and efficiency are substantial and hard to dispute \cite{TonerRodgers2024}.

While we may marvel at the productive outcomes of human and LLMs collaboration in business, it is perhaps worth questioning its impact on the education sector. Indeed, although delegating cognitive tasks to AI may pose little issue for trained adults in a professional work setting, the reliance of young learners on AI for key tasks —such as critical thinking, summarisation, and even basic logic— should raise some concern. LLMs, and particularly ChatGPT, are widely used by all demographics, including children and young adults, who, according to educational professionals, are substantial users of these tools in schools and universities. On one hand, controlled and guided use of LLMs for educational purposes could significantly enhance learning by enabling customised programs and teaching AI assistants \cite{baskara2023exploring,su2023collaborating,lee2022developing} and could be a helpful technology for students with learning disabilities \cite{ayala2023chatgpt,tamdjidi2023chatgpt}. On the other hand, although AI brings opportunities for education, it may also present risks \cite{rahman2023chatgpt}: the unsupervised and extensive use of these tools, for example for home assignments that were designed before the public release of AI assistants, may lead to unintended consequences. Such outcomes could include the decline of essential cognitive skills or a weaker foundation in basic knowledge, both of which are crucial for mastering more complex concepts and subjects.

In this context, we conducted a survey targeting secondary school, high school, and university students aged 13 to 25 in France and Italy, aiming to gain a clearer understanding of ``how much''and ``how'' they utilise LLMs tools in educational settings. Our study seeks to quantify the use of these AI tools across student groups sorted by age and gender, identify any trends indicating specific subjects where these tools are most frequently applied to, and assess students' ability to critically evaluate the limitations of LLMs across demographic categories. From this, we aim to detect early indicators of a potential emerging \emph{AI gap} \cite{bentley2024digital}, a divide that could mirror the existing \emph{digital gap} \cite{herbert2017digital,longoria2022systematic} or \emph{connectivity divide}. We will also offer preliminary insights into the potential decline of certain cognitive skills, such as critical thinking and writing. Finally, we hope to provide suggestions for education professionals on adapting teaching strategies to leverage these AI innovations effectively.

Summarizing, we investigate three main research questions:
\begin{itemize}
    \item How common is the use of LLMs for academic purposes among student populations (from 13 years old)?
    \item Can we detect trends and differences depending on the age group, the gender, or even topics?
    \item Based on the answers to the previous two questions, what issues --already existing or new-- can we foresee? And what can we recommend to education professional to mitigate them?  
\end{itemize}

This paper is organized as follows: in Section~\ref{sec:sota} we present some state of the art on LLMs and the impact of new technologies on education. Section~\ref{sec:material} presents the methodology we applied: the questionary we proposed, the demographic of our respondents and the statistical tools used to interpret the results. Section~\ref{sec:analysis} is concerned with the presentation of the results and their interpretation. Finally, the paper ends with a conclusion and a discussion about future perspectives.

\section{State of the art}
\label{sec:sota}
The 21st century is often regarded as the era of technology, which plays a vital role in economic growth and efficiency. In education, technology has transformed how students learn, making education more interactive, efficient, and accessible. Modern tools like the internet have significantly enhanced learning by providing continuous connectivity, access to tutorials, and interactive visual aids. Students can now easily find educational resources online, which improves their understanding and engagement. The integration of digital media has further revolutionized the education sector, enabling round-the-clock support and access to various learning platforms. Additionally, the rise of online degrees has made education more flexible and accessible, allowing students to earn certifications and degrees through digital platforms. This technological shift in education is continually growing and reshaping the way students learn and engage with academic material.

\subsection{New Technologies and Education}\label{subsec:new_tec}
As underlined by \citet{raja2018impact}, Information and communications technologies (ICT) play a vital role in education in four key ways: as an integral part of the curriculum, as a tool for instructional delivery, as a means of supporting teaching, and as a tool to enhance the overall learning experience. Thanks to technological advancements, education has shifted from being passive and reactive to interactive and engaging \cite{ben1998constructivism,sentance2017creating,fessard2019there}. In both corporate and academic environments, education serves different purposes: training employees to improve performance in the workplace, and fostering curiosity and critical thinking in students. In both contexts, technology helps learners grasp and retain concepts more effectively. In this context, teachers face lots of challenges due to the rapid expansion of knowledge and the increasing role of technology in education. In fact, they are required to adapt to new technologies, which intensifies their training needs. \citet{gressard1985} emphasize that teachers' attitude toward computers is crucial for successful ICT integration, noting that negative attitudes can hinder the success of computer-based initiatives. Common barriers to technology adoption include lack of time, access, resources, expertise, and support. 
Additionally, reliability issues such as hardware failures, incompatible software, slow internet connectivity, and outdated software at schools, compared to more up-to-date software at home, also pose challenges, as noted by 
\citet{butler2002} and \citet{chizmar2001}.

 According to \citet{tinio2020}, ICT significantly impacts the acquisition and absorption of knowledge for both teachers and students by promoting various learning approaches such as ``active learning'', ``collaborative learning'', ``creative learning'', ``integrative learning'' and ``evaluative learning''. Overall, ICT enhances the learning experience by fostering engagement, collaboration, creativity, and critical thinking. All this led to both positive and negative aspects. On the one hand, technology in education enables a more exciting and engaging learning experience for students; it provides flexibility for those with busy schedules, allowing them to work at their own pace and from home; additionally, it helps students develop valuable technology skills that will benefit them in the workplace and it reduces the need for paper and photocopying, contributing to environmental sustainability. On the other hand, there are also some disadvantages to take into account: for instance, some experts argue that the widespread use of technology in education can hinder students' imagination \cite{james2012}, and reduce their critical thinking abilities \cite{mayer2020}.  
 Furthermore, some critics highlight the adverse health effects associated with prolonged screen use. Indeed, various studies have shown that excessive screen time and media multitasking can negatively impact executive functioning, sensorimotor development, and academic performance \cite{muppalla2023effects,tadpatrikar2024digital}. Furthermore, from the teacher's perspective, integrating technology can be time consuming and can lead to difficulties in evaluating the actual improvement in students' knowledge. 

We can therefore surely assert that technologies have profoundly transformed the educational landscape. In particular in recent years Computer Science (CS), encompassing fields such as Information and Communication Technologies (ICT) and Artificial Intelligence (AI), have been pivotal in driving this transformation. Numerous reports have extensively documented the impact of these technologies on education \cite{al2024technology,ccelik2024technology}. Below, we briefly sum up some rising technologies in education to show the critical role that computer science plays in and its potentials :
\begin{enumerate}
    \item Information and Communication Technology (ICT)
\begin{itemize}
    \item \textbf{Computers and the internet} provide students with access to extensive educational resources, including digital textbooks, instructional videos, and online materials, enabling learning beyond the classroom;
    \item \textbf{Online learning platforms} like Moodle, Google Classroom, and Canvas facilitate course administration, assignment management, and student communication in schools and universities;
    \item \textbf{E-books and digital materials} provide students with easy and cost-effective access to academic resources like textbooks and journals;
    \item \textbf{Webinars and video conferences} enable students to participate in virtual lectures and seminars from any location.
\end{itemize}
\item Artificial Intelligence (AI)
\begin{itemize}
\item \textbf{Personalized Learning}: AI helps identify each student's unique needs and delivers tailored learning content based on their skill level and interests;
\item \textbf{Student Performance Measurement}: AI enables rapid assessment of student performance on a large scale, allowing teachers and tutors to provide timely feedback;
\item \textbf{Educational chatbots} use AI to offer students instant help with questions and curriculum-related information.
\end{itemize}
\item Virtual Reality (VR) and Augmented Reality (AR)
\begin{itemize}
    \item they enhance learning by creating immersive experiences, allowing students to explore historical sites, scientific concepts, and realistic simulations interactively;
    \item they are also utilized in teacher training to offer practical experience in managing diverse classroom scenarios.
\end{itemize}
\item Game-Based Online Learning
\begin{itemize}
    \item \textbf{Serious Games} are specifically designed for educational purposes \cite{breuer2010so} and can be used to teach complex concepts in an engaging and interactive way;
    \item The \textbf{gamification} approach incorporates game elements like points, levels, and rewards to enhance student motivation and engagement in the learning process.
\end{itemize}
\item Education Analytics
\begin{itemize}
    \item \textbf{Data mining and analytics} in education help schools and colleges identify student performance patterns, predict dropouts, and analyze learning trends;
    \item \textbf{Learning analytics} involves measuring, collecting, analyzing, and reporting data about learners and their context to understand and enhance both learning and the environment in which it takes place.
\end{itemize}
\item Robots
\begin{itemize}
    \item Robotics in education is an emerging field where robots teach students various subjects, interacting with them using human-like facial expressions and emotion-detection technology. Robots serve in roles such as teaching assistants, personal tutors, small group leaders, and peer learners. While they are widely used in STEM education, robots are also effective in teaching humanities subjects, including language learning.
    \item they are particularly exploited  to enhance learning outcomes and social skills for students with autism and special needs. 
\end{itemize}
\end{enumerate}

\subsection{Large Language Models}\label{subsec:llms}

Language modeling has been a well-established field of research since the 1950s, when C.E. Shannon first applied information theory to human language \cite{shannon1950}. Shannon's groundbreaking work introduced the idea of using statistical models to predict and compress natural language text, laying the foundation for the first wave of language modeling with n-gram models. Over the years, the field has progressed through four distinct waves \cite{minaee2024}, each making significant contributions to the development of modern Natural Language Processing (NLP) systems: Statistical Language Models (SLMs), Neural Language Models (NLMs), Pre-trained Language Models (PLMs) and Large Language Models (LLMs).
SLMs \cite{zhai2008} modeled text as a sequence of words, predicting each word's probability based on previous ones. N-gram models, using Markov chains, were commonly used in various NLP tasks but faced issues with data sparsity and required smoothing techniques for unseen words or sequences \cite{CHEN1999359}.
NLMs addressed the limitations of SLMs by mapping words into low-dimensional continuous vector spaces, known as word embeddings: neural networks were used to predict the next word in a sequence by aggregating the embeddings of preceding words; this innovation alleviated data sparsity issues also allowing for the computation of semantic similarity between different linguistic inputs \cite{bengio2000neural}. However, early NLMs were primarily task-specific, limiting their general applicability.
Conversely, PLMs are task-agnostic; they (as BERT \cite{devlin2018bert} and GPT \cite{radford2018improving}) introduced a paradigm shift through the pre-training and fine-tuning process where language models, based on recurrent neural networks or transformers, are pre-trained on large, unlabeled text corpora for general tasks like word prediction, and then fine-tuned on small, labeled datasets for specific tasks \cite{zhou2024comprehensive}.
LLMs, exemplified by models like LLaMA \cite{grattafiori2024llama}, Mixtral \cite{jiang2024mixtralexperts}, Gemma\cite{gemmateam2024gemmaopenmodelsbased}, Qwen2 \cite{yang2024qwen2technicalreport}, PaLM\cite{chowdhery2022palmscalinglanguagemodeling} and GPT-4 \cite{openai2024gpt4technicalreport},  are transformer-based models, containing billions of parameters, which are trained on vast text corpora and exhibit advanced language understanding and generation capabilities; LLMs demonstrate emergent abilities such as in-context learning (being capable of any downstream tasks without any gradient update or fine-tuning), instruction following, and multi-step reasoning, which were absent in smaller models like PLMs. Thanks to these abilities and the augmentation through external tools and continual learning mechanisms, LLMs can be seen as foundational components in the development of Artificial General Intelligence (AGI).

LLMs, now equipped with billions of parameters, have consistently shown that increasing model size correlates with enhanced capabilities: this trend is further supported by the growth in the scale of training datasets, which now reach approximately more than 1 trillion tokens (an order of magnitude larger than previous models). For example, GPT-2 was trained on 10 billion tokens, while GPT-4 was trained on approximately 14 trillion, leading to significant improvements in performance \cite{minaee2024}.

A pivotal moment in the mainstream adoption of LLMs occurred in November 2022 with the release of \textit{ChatGPT} \cite{openai2023chatgpt} as this development allowed the general public to interact with these sophisticated models at minimal cost, thereby democratizing access to cutting-edge AI. As a result, it catalyzed a widespread perception that artificial intelligence had made a substantial leap forward, shifting the narrative from futuristic speculation to present-day reality.

Central to the success of LLMs is the self-attention mechanism, which enables these models to generate coherent and contextually relevant text by capturing long-range dependencies between words. By attending to the most relevant parts of the input sequence, LLMs are able to model complex linguistic structures, resulting in highly accurate and contextually appropriate outputs. This capacity for nuanced understanding and generation has made LLMs essential tools across a wide array of natural language processing tasks \cite{zhao2024surveylargelanguagemodels}.

\subsection{ChatGPT and Education}

Considering the context outlined in SubSection \ref{subsec:new_tec} and given the rapid surge of LLMs, as discussed in \ref{subsec:llms}, as well as their widespread adoption by the general public, particularly exemplified by ChatGPT, it is natural to ask how these tools might transform the way people learn and what their short- and long-term impacts could be in the learning processes and in education more generally.

On one hand, these tools could be seen as helpful \cite{Ali_Shamsan_Hezam_Mohammed_2023,kasneci2023chatgpt,dong2024}, also for supporting students with difficulties \cite{ayala2023chatgpt,tamdjidi2023chatgpt}. On the other hand, there is a concern that students might rely on them to solve any type of school task, drastically undermining their learning process \cite{ratnam2023chatgpt,memarian2023chatgpt} and the potential of false information as well as compromised
academic integrity
\cite{farhi2023analyzing}. These are some concerns \cite{meyer2023chatgpt,tlili2023if,bhullar2024chatgpt} that have been investigated in recent researcher works. 

The review of \cite{albadarin2024systematic} on 14 selected empirical studies  on ChatGPT from students and teachers points of views underline these two opposite sides: 
on the positive one, ChatGPT significantly supported the learning process in various ways; learners utilized it as a virtual intelligent assistant, benefiting from immediate feedback, on-demand answers, and easy access to educational resources. It was particularly effective in improving writing and language skills, helping learners generate ideas, compose essays, summarize, translate, paraphrase texts, and check grammar. Additionally, ChatGPT facilitated personalized and directed learning, assisting with understanding concepts, completing homework, creating structured learning plans, and clarifying assignments.
Educators also found ChatGPT valuable for boosting productivity and efficiency. It was used to create lesson plans, design quizzes, provide additional resources, and answer students’ questions, saving time and enabling more dynamic and engaging teaching strategies.
On the other hand, the findings indicated that excessive reliance on ChatGPT could diminish learners' creativity and collaborative learning. Over-dependence on the tool for quick answers may hinder critical thinking and problem-solving skills, as students may avoid engaging deeply with the material or exploring alternative solutions. This issue was particularly noticeable in group projects, where students often turned to ChatGPT individually instead of brainstorming and collaborating with peers, negatively impacting teamwork. Additionally, the integration of ChatGPT into education has raised broader concerns, such as the risk of inaccurate or misleading information, inequitable access, academic integrity challenges, and potential misuse of the technology.

\citet{punar2024cultivating} investigated ChatGPT's impact on writing tasks, focusing on two main issues: on one side, the potential of ChatGPT as a learning assistant to enhance students' self-editing skills in writing; on the other, students' opinions and recommendations regarding the use of ChatGPT as a learning assistant. They found that ChatGPT effectively supports students in developing formal writing skills by providing valuable suggestions and corrections; however, challenges included technical issues and students highlighted the need for functional improvements to make ChatGPT more effective for self-editing; despite its potential, the study’s findings were limited by a small sample size of 11 participants, which affects the generalizability of the results.

\citet{mogavi2024chatgpt} investigated the adoption and perception of ChatGPT in education by analyzing qualitative data collected from various social media platforms; with the aim of  understand the user experience and views of early adopters of ChatGPT across different educational sectors, their analysis revealed that ChatGPT is utilized in diverse settings for multiple purposes, with the most widespread uses observed in higher education, K-12 education, and practical skills development. 

A survey \cite{stojanov2024university} was conducted on 490 university students recruited via CloudResearch to  examine their reliance on ChatGPT for completing 13 general tasks related to learning processes. Firstly the survey seek to characterize the nature of students' reliance on ChatGPT; secondly, it aims to understand the relationship of ChatGPT reliance profiles to students' AI literacy, their attitudes towards ChatGPT, their critical usage of AI tools, and students’ achievement goal orientations. The study identified five distinct profiles: Versatile Low Reliers (38.2\%) showed low overall reliance on ChatGPT for tasks, while All-Rounders (10.4\%) reported high reliance across the board; Knowledge Seekers (16.5\%) heavily used ChatGPT for content acquisition, information retrieval, and text summarization, whereas Proactive Learners (11.8\%) relied on it for feedback, planning, and quizzing. Lastly, Assignment Delegators (23.1\%) used ChatGPT extensively for drafting assignments, completing homework, and even having it write assignments. These findings suggested that while ChatGPT may not be the primary learning aid for many students, it plays a distinctive role in their learning.

Another interesting research \cite{al2024understanding} investigates the perceptions of GenAI among 366 students in the United Arab Emirates using survey data. Factor analysis identified relevant scales, followed by mean comparisons based on students’ agreement with a statement about their willingness to use ChatGPT. The findings reveal high awareness among participants of ChatGPT's benefits, limitations, and risks. Awareness of benefits was linked to a stronger intention to use ChatGPT in the future. Surprisingly, awareness of limitations and risks also positively correlated with the willingness to use ChatGPT, challenging the notion that these factors act solely as barriers and highlighting a complex relationship between risk perception and technology adoption in education. Another study \cite{qu2024disciplinary}, surveying 193 undergraduate students from a university in Singapore, examines their knowledge, usage intentions, and engagement with Generative AI (GenAI) across academic disciplines using a hard/soft and pure/applied framework.  They used a Google Forms questionnaire with closed-ended, open-ended, and Likert scale questions to gather both quantitative and qualitative data concerning demographic characteristics (assessed students' study level, academic discipline, and GenAI awareness and knowledges) and GenAI Usage (explored the types of GenAI tools used, academic tasks involving GenAI, and reasons for usage or non-usage). Their findings reveal significant disciplinary differences: applied fields (both hard and soft) demonstrate higher levels of GenAI knowledge and usage intentions compared to pure fields. While engagement with GenAI for routine tasks is consistent across disciplines, engagement in cognitive tasks is notably higher in applied fields. These results highlight how the practical focus of applied fields drives GenAI adoption in academic contexts.

Other students surveys articles, more similar to our approach, have primarily focused on the factors driving students to use ChatGPT. Among these, a Peruvian study \cite{velasquez2024main} targeted undergraduate students and identified writing capabilities, the variety of information, perceived usefulness, and ethical considerations as the main factors motivating the use of LLMs. A similar study in Oman \cite{tiwari2024drives}, which surveyed a larger sample of undergraduate and graduate students, extended this focus. Beyond examining behavioural motivations, the Omani study highlighted the challenges associated with the rapid and uncontrolled proliferation of LLMs in academia, including bias amplification, ethical concerns, a lack of in-depth comprehension, and a general decline in critical thinking skills attributed to these AI tools.

Lastly, two other comparable studies conducted in Malaysia (406 higher education students) \cite{foroughi2024determinants} and Poland (534 higher education students) \cite{strzelecki2023use} reached similar conclusions, further reinforcing these trends. It is worth noting that, while our methodology for collecting and evaluating survey results differs and will be discussed in the relevant section, our study further diverges from these works in its focus. Rather than examining the factors motivating students to use ChatGPT, we aim at investigating the extent and variations in usage across different student populations, segmented by age, gender, and academic disciplines. Furthermore, our survey encompasses a broader age range, including minors, offering a more comprehensive perspective with the aim of having a more general idea of ChatGPT use by students and of proposing some general and possible recommendation for changing in educational settings.

\section{Material and Methods}
\label{sec:material}

In this section, we begin by describing the main characteristics of our respondents population, the content of our survey. Then, we present the statistical tools that we will use in Section \ref{sec:analysis} to interpret the survey results and draw our conclusions.

\subsection{Survey content}

Our Google Form survey was distributed to professors and heads of institutions to facilitate its administration directly in classrooms via a one page document in which the general aim of the survey was presented. This document featured a QR code link to the survey's form webpage, enabling respondents to complete it using their mobile phones as well as a direct link for those who preferred to respond via a personal computer or laptop. We included a statement about the complete anonymity of the results, and the guaranty of processing the answers in a global way and not individually in accordance with the General Data Protection Regulation\footnote{(GDPR - EU Regulation 2016/679 and Legislative Decree no. 196/2003)}. For students under the age of 16 and requiring parental authorization before participating, a one-page document was provided to their parents outlining the survey's objectives and including essential legal information, such as the EU GDPR notice. Professors responsible for administering the survey during class time were instructed to read the one-page document aloud and display it (for instance using a projector) to let the students access the questions with the QR code or the link.

The survey itself comprised the following question items:
\begin{enumerate}
    \item \textit{How old are you ?}
    \item \textit{How do you identify yourself ?} With the possibility of skipping the question, or entering an alternative gender other than male or female.
    \item \textit{Have you used ChatGPT (or an equivalent program) at least once during the course of your studies or in an academic context ?} [Yes | No]. In case of a negative answer, the survey skipped to Item 8.
    \item A set of 4 questions detailing their use of ChatGPT for humanity related topics (literature, history, geography, social sciences, foreign languages, law, redaction, etc.):
        \begin{itemize}
            \item \textit{Have you ever used ChatGPT for a ``humanities'' topic ?} [Never | At least once | A few times | Often].
            \item \textit{What did you think of the results ?} [Does not apply. I did not use it | It was very bad | It was mediocre | It was quite good | It was very good].
            \item \textit{If you used it, did you rework the answers provided by ChatGPT ?} [Does not apply. I did not use it | No, never | Yes, sometimes | Yes, always].
            \item \textit{For which of the following ``humanities'' topics or purpose did you use ChatGPT? Check all that apply.} For this question, we listed a list of common humanity fields and left the possibility of adding some. 
        \end{itemize}
    \item A set of 4 questions detailing their use of ChatGPT for scientific related topics (mathematics, physics, chemistry, biology, computer science etc.):
        \begin{itemize}
            \item \textit{Have you ever used ChatGPT for a technical or science topic ?} [Never | At least once | A few times | Often].
            \item \textit{What did you think of the results ?} [Does not apply. I did not use it | It was very bad | It was mediocre | It was quite good | It was very good].
            \item \textit{If you used it, did you rework the answers provided by ChatGPT ?} [Does not apply. I did not use it | No, never | Yes, sometimes | Yes, always].
            \item \textit{For which of the following technical/science topics or purpose did you use ChatGPT? Check all that apply.} For this question, we listed a list of common science fields and left the possibility of adding some. 
        \end{itemize} 
    \item \textit{Which of the following devices do you use ChatGPT with ?} [PC or laptop | smartphone | tablet | Voice recognition with any of the previous ones]. Multiple choices where possible.  
    \item \textit{Have you ever used ChatGPT outside an academic context ?} [Never | At Least once | sometimes | Often]. If the answer was ``never'', the survey stopped there.
    \item \textit{If you used ChatGPT in a non-academic context, please specify in which contexts you used it:}  [multiple choices - open text answer] (see Section~\ref{subsec:nonusers}). The survey stopped here for all respondents unless for the one who answered ``No'' to Item 3, which were directly sent to Item 9.
    \item \textit{Why have you never used ChatGPT ?} Multiple choices: [I did not know this tool | I know this tool, but I did not think it was useful | I know this tool, but I never thought of using it this way | I do not think that it is honest to use ChatGPT in an academic setting | Other]. The ``other'' option was left with the possibility of writing a short text.
\end{enumerate}

\subsection{Respondents demographics}

Our survey targeted students aged 13 to 25 years old. It was conducted in France and Italy, where various high schools and higher education institutions agreed to allow some of their classes to participate in our study. Unlike previous surveys conducted online without supervision \cite{tiwari2024drives,velasquez2024main,foroughi2024determinants}, ours was administered during class time under the supervision of a professor. Moreover differently from  \citet{stojanov2024university} and \citet{qu2024disciplinary} responders were volunteers and they not perceived any kind of financial reward. This approach also enabled us to collect data from younger age groups compared to other studies.

The survey was conducted between May 2024 and October 2024, yielding a total of 395 responses. The distribution of respondents by age and gender is presented in Table \ref{tab:global}.

\begin{table}[!h]
\centering
\begin{tabular}{lrr}
 \toprule
 \textbf{Category} & \textbf{\emph{n}} & \textbf{\emph{\%}} \\
\midrule
 Male & 211 & $53.4$\% \\
 Female & 174 & $44.1$\% \\ 
 Other & 9 & $2.3$\% \\
 No answer & 1 & $0.3$\% \\
\midrule
13--16yo & 99 & $25.1$\% \\
17--19yo & 107 & $27.1$\% \\
20--22yo & 152 & $38.5$\% \\
23+yo & 37 & $9.4$\% \\
\midrule
English form & 99 & $25.1$\% \\
French form & 200 & $50.6$\% \\
Italian form & 96 & $24.3$\% \\
\midrule
 Used ChatGPT & 344 & $87.1$\% \\
 Never used ChatGPT & 51 & $12.9$\% \\ 
\bottomrule
\end{tabular}    
\caption{Overview of our respondent population (N=395).}
\label{tab:global}
\end{table}

For presentation purposes, we categorised respondents into age groups corresponding approximately to secondary school (13 to 16 years), high school (17 to 19 years), undergraduate studies (20 to 22 years), and graduate studies (23 years and older).

Given the young age of some respondents and the international composition of undergraduate and graduate student populations, different versions of the survey were provided in English, French, and Italian. This approach ensured that all respondents could complete the survey in a language they were proficient in.

However, the language of the survey cannot be used to infer the respondents' nationality or native language, for the following reasons:
\begin{itemize}
\item A majority of Italian higher education students opted to complete the English version of the survey.
\item Many French higher education students answered in French, including some international students.
\item Two high school classes surveyed were international classes taught in English, and some students chose to answer in English rather than French or Italian.
\end{itemize}

It is worth noting that, despite responding positively to question item 3, a few students indicated that they had never used ChatGPT for both items 4.1 and 5.1. This inconsistency leads to discrepancies, with some students being classified as non-users of ChatGPT in parts of subsections \ref{sec:age} and \ref{sec:gender} (see Table \ref{tab:age_fu}).

Finally, we acknowledge a probable bias in the undergraduate and graduate populations surveyed. Unlike high school, where a broad range of subjects is taught, university respondents were predominantly from science-related fields, as students in humanities-related fields proved more difficult to reach.

\subsection{Statistical tools used to interpret the answers}

To analyse and interpret the results of our survey, we will use standard visualisations such as bar plots, histograms and pie charts, as well as statistical tools such as chi-squared test analysis and confidence intervals.

For completeness, we remind readers that the chi-squared test is a statistical hypothesis test commonly employed to analyse contingency tables. Its primary objective is to assess whether a significant association exists between categorical variables. Specifically, it evaluates whether a statistically significant difference is present between the expected frequencies, $e_{i,j}$ (assuming a perfectly uniform distribution across categories), and the observed frequencies, $o_{i,j}$, in one or more categories of a contingency table with $r$ rows and $c$ columns, as described in Equation \eqref{eq:chi2}:

\begin{equation}
	\chi^2 = \sum_{i=1}^r \sum_{j=1}^c \frac{(o_{i,j} - e_{i,j})^2}{e_{i,j}}
\label{eq:chi2}
\end{equation}	

The result of a chi-squared test is typically analyzed using the following steps:
\begin{itemize}
    \item First, the p-value of the test must be computed an analyzed. Typical cut-off values to reject the null hypothesis for the p-value are $0.01$, $0.05$ or $0.1$ to reject the independence hypothesis with a confidence of $99\%$, $95\%$ or $90\%$ respectively.
    \item Then, if the independence hypothesis is rejected and an association between certain categories is therefore confirmed to be statistically valid, it is common practice to look at the normalized residuals between the expected and observed values in order to detect which pairs of categorical variables are under-represented or over-represented (see Equation \eqref{eq:res}).

\begin{equation}
	r_{ij} = \frac{o_{ij} - e_{ij}}{\sqrt{e_{ij}}}
\label{eq:res}
\end{equation}	
    
    \item It is also possible to analyse the contribution to the chi-squared in order too detect the pairs that most contributed to build a significant statistical relationship (see Equation \eqref{eq:contr}).

\begin{equation}
	c_{ij} = \frac{r_{ij}^2}{\chi^2}
\label{eq:contr}
\end{equation}	
    
    \item Finally, although it is mostly irrelevant for this study, it is possible to compute the relative strength of the statistical relationship using indexes such as the Cramer index.
\end{itemize}

Lastly, some parts of our study may refer to confidence intervals computed under the binomial-normal law approximation. Given a sample of size $N$, this is a commonly used method to assess the confidence interval of an outcome of estimated probability $\hat{p}$ versus all other possible categories and outcomes. It is computed as shown in Equation \ref{eq:ic95bin}, where unless stated otherwise we will use $L^N_\alpha=1.96$ which is the standard value for a 95\% confidence interval under the normal law hypothesis. 

\begin{equation}
   p = \hat{p} \pm L^N_\alpha \sqrt{ \frac{\hat{p}(1-\hat{p})}{N}}
   \label{eq:ic95bin}
\end{equation}

\section{Survey answers analysis}
\label{sec:analysis}

For a better understanding of this section, we first remark that the number of respondents may vary from one table to another depending on the different sub-studies. The reasons are the following:
\begin{itemize}
    \item10 respondents did not specify their gender and have been discarded from the parts focusing on gender differences;
    \item while 346 out of 395 respondents indicated that they were ChatGPT users in question 3 of our survey, 8 of them declared that they never used in to both questions item 4.1 and 5.1, thus making them de facto non-users in several sub-studies. Two of them also belong to the previous category that did not declare their gender.
\end{itemize}

For these reasons, depending on the subject and exclusion criteria of the sub-study, we considered 328 to 395 respondents.


\subsection{Age trends analysis}
\label{sec:age}

In this subsection, we analyse the results of our survey, focusing on potential differences in the academic use of ChatGPT and other LLMs across the age groups of our respondents. As a reminder, the overall age distribution is provided in Table \ref{tab:global}.

\begin{figure}[!h]
    \centering
    \includegraphics[width=0.66\textwidth]{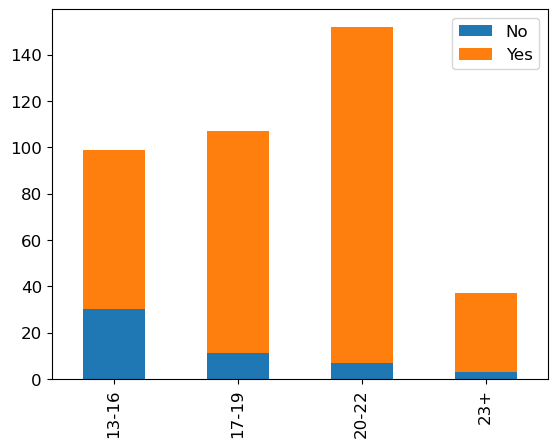}
    \caption{ChatGPT use by age range based on answers to question item number 3.}
    \label{fig:age_use0}
\end{figure}

The first notable finding from our study is the widespread use of ChatGPT and LLMs in academia, as reported by students. As illustrated in Figure \ref{fig:age_use0} derived from Table \ref{tab:global}, even among the youngest respondents, aged 13 to 16, nearly 70\% indicated that they had used such tools at least once in an educational context. Secondary school students stand out in particular, accounting for 71\% of the chi-squared value, with a p-value below $10^{-7}$.

Among other age groups, the use of ChatGPT increases significantly, ranging from 89.7\% to 95.4\%, with students aged 20 to 22 emerging as the most active users. 

\begin{figure}[!h]
    \centering
    \includegraphics[width=0.66\textwidth]{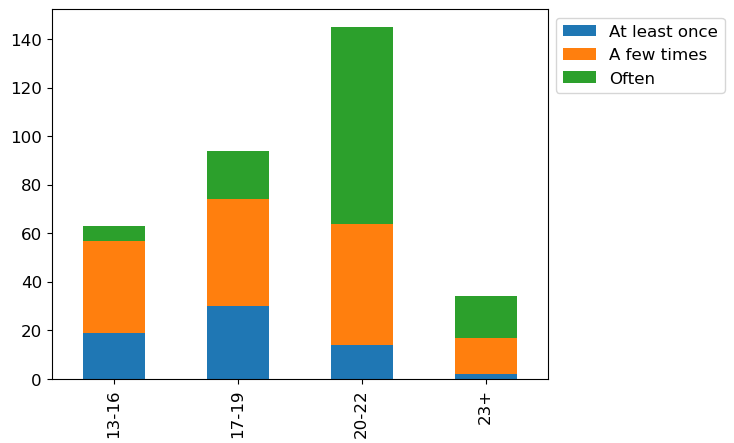}
    \caption{ChatGPT use frequency by age range based on answers to question items 4.1 and 5.1, excluding non-users.}
    \label{fig:age_use1}
\end{figure}

By combining items 4.1 and 5.1 (namely students's ChatGPT use in humanities and scientific topics), and excluding students who do not use LLMs, we can focus on the frequency of use among students familiar with ChatGPT. The results, shown in Figure \ref{fig:age_use1}, are derived directly from Table \ref{tab:age_fu}.

\begin{table}[!h]
\centering
\begin{tabular}{lrrrr}
 \toprule
 & 13-16 & 17-19 & 20-22 & 23+ \\
\midrule
At least once & 19 & 30 & 14 & 2 \\
A few times & 38 & 44 & 50 & 15 \\
Often & 6 & 20 & 81 & 17 \\
\midrule
Non-Users & 30 (+6) & 11 (+2) & 7 & 3 \\
\bottomrule
\end{tabular}    
\caption{Age repartition and frequency of use (N=395).}
\label{tab:age_fu}
\end{table}

From Figure \ref{fig:age_use1}, a trend similar to that observed in Figure \ref{fig:age_use0} emerges, with the frequency of use increasing with age. The chi-squared test yields a p-value of $9 \times 10^{-12}$ ($\chi^2=63.43$), and further residual and chi-squared contribution analysis (see Table \ref{tab:age_fur} for details) reveals the following:
\begin{itemize}
    \item Students aged 13 to 16 years form a distinct group, using ChatGPT significantly less frequently than older groups. This is reflected in negative residuals for the most frequent uses (``Often'') and a chi-squared contribution of 20\% for this pairing.
    \item Conversely, students aged 20 to 22 years stand out as the most avid users of LLMs, as indicated by positive residuals for ``Often'' and a chi-squared contribution of 22\% for the same pairing.
\end{itemize}

\begin{table}[!h]
\centering
\begin{tabular}{lrrrr}
 \toprule
 & 13-16 & 17-19 & 20-22 & 23+ \\
\midrule
At least once & 2 (6\%) & 2.8 (12\%) & -2.7 (11\%) & -1.2 (5\%) \\
A few times & 2 (6\%) & 0.45 (0\%) & -1.7 (4\%) & 0.032 (0\%) \\
Often & -3.6 (20\%) & -2.5 (10\%) & 3.8 (22\%) & 1.3 (2\%) \\
\bottomrule
\end{tabular}    
\caption{Normalized residuals and contribution to the chi squared (value between parenthesis) for the age repartition and frequency of use, excluding non-users (N=336).}
\label{tab:age_fur}
\end{table}

\begin{figure}[!h]
    \centering
    \includegraphics[width=0.66\textwidth]{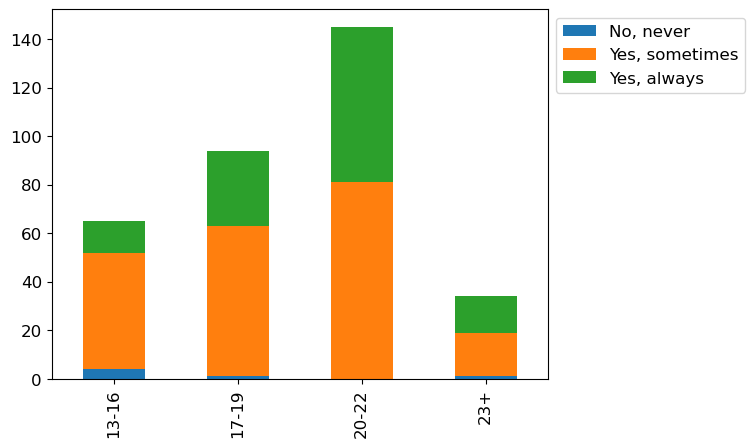}
    \caption{User behavior regarding the reworking of ChatGPT answers by age range based on questions items 4.3 and 5.3, excluding non-users.}
    \label{fig:age_rework}
\end{figure}

Focusing on the habit of reworking ChatGPT-generated responses, using items 4.3 and 5.3 and excluding non-users, we observe the results in Figure \ref{fig:age_rework}, derived from Table \ref{tab:age_rw}. The proportion of students reporting that they never proofread or rework LLM-generated answers is minimal, with 6.2\% being the highest figure among younger demographics. While there may be a bias due to the survey context, it is clear that students across all age groups recognise the importance of revising LLM-generated outputs. In particular, it is worth to underline that the proportion of students who systematically rework these responses increases with age. This result could be correlated to the fact that the ``older'' group is more accustomed to the tool and therefore knows better how it works and what limits they can expect from it. 

However, even among the oldest students, systematic proofreading remains below 50\%. Among students aged 13 to 16, only 20\% declared to systematically revise answers, and this group accounts for a total chi-squared contribution of 65\% ($\chi^2=21.22$, p-value = $0.0017$), highlighting their distinct behaviour compared to other age groups.

\begin{table}[!h]
\centering
\begin{tabular}{lrrrr}
 \toprule
 & 13-16 & 17-19 & 20-22 & 23+ \\
\midrule
No, never & 4 & 1 & 0 & 1 \\
Yes, sometimes & 48 & 62 & 81 & 18 \\
Yes, always & 13 & 31 & 64 & 15 \\
\midrule
Non-Users & 30 (+6) & 11 (+2) & 7 & 3 \\
\bottomrule
\end{tabular}    
\caption{Age repartition and reworking the answers (N=395).}
\label{tab:age_rw}
\end{table}

\begin{figure}[!h]
    \centering
    \includegraphics[width=0.66\textwidth]{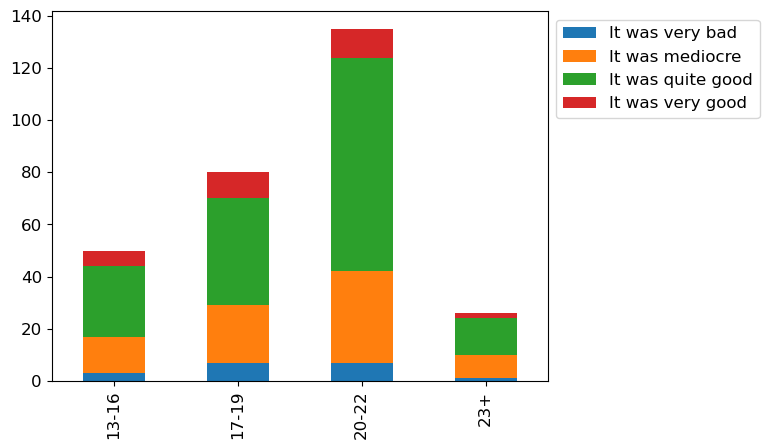}
    \caption{Satisfaction of ChatGPT answers by age groups based on questions items 4.2 and 5.2, excluding non-users.}
    \label{fig:age_sat}
\end{figure}

We now turn to the satisfaction analysis of LLM-generated answers by age group, shown in Figure \ref{fig:age_sat} and derived from Table \ref{tab:age_sat}. The results indicate a homogeneous situation across all age groups, as confirmed by a chi-squared test p-value of $0.90$ ($\chi^2=4.12$). Overall satisfaction is high, with the option ``It was quite good" representing more than 50\% of responses in all groups. Cases of very high satisfaction (``It was very good") or dissatisfaction are rare, though the proportion of ``It was very good" responses tends to decrease with age.

Finally, these results may also be interpreted as showing that LLMs do not appear to become less relevant or efficient even at higher curriculum levels.

As an initial conclusion to this sub-study by age group, we identify the following key findings:
\begin{enumerate}
    \item While we had clues that LLMs were used by students as young as 13 to 16 years old, our results show that despite somewhat lower usages compared with university students, LLMs are already massively used by this category of students. Currently, some universities have started thinking about adapting their curricula or introduced sessions to address these tools, however it is evident that such initiatives should begin much earlier, ideally in secondary school.
    \item The extensive use of LLMs and high satisfaction levels across all age groups make it clear that ignoring these tools is no longer a viable option. This raises questions about the effectiveness of traditional homework as a learning and practice tool, given the likelihood of AI assistance in its completion.
    \item On a positive note, nearly all students who use these tools are aware of the need to revise LLM-generated answers. However, this practice is less systematic among younger students, suggesting an area for targeted educational improvement.
\end{enumerate}

\begin{table}[!h]
\centering
\begin{tabular}{lrrrr}
 \toprule
 & 13-16 & 17-19 & 20-22 & 23+ \\
\midrule
It was very bad & $6\%$ & $8.7\%$ & $5.2\%$ & $3.8\%$ \\
It was mediocre & $28\%$ & $27.5\%$ & $25.9\%$ & $34.6\%$ \\
It was quite good & $54\%$ & $51.2\%$ & $60.7\%$ & $53.8\%$ \\
It was very good & $12\%$ & $12.5\%$ & $8.1\%$ & $7.7\%$ \\
\bottomrule
\end{tabular}    
\caption{Average satisfaction based on the age, excluding non-users (N=336).}
\label{tab:age_sat}
\end{table}

These results should encourage a push for re-thinking homeworks, and in particular designing them in a way that accounts for ChatGPT as a potential support. As such, perhaps the emphasize should be shifted to subject appropriation, fundamental understanding, critical thinking and memorization rather than raw answers which can and will be delegated in part or in full to LLMs.

\subsection{Gender trends analysis}
\label{sec:gender}

In this subsection, we analyse the results of our survey, focusing on potential differences in the academic use of ChatGPT and other LLMs by gender. As a reminder, the gender distribution is presented in Table \ref{tab:global}. Note that due to the low number of respondents in these categories, we excluded the 10 students who either did not answer the gender question or identified as other than male or female.

\begin{figure}[!h]
    \centering
    \includegraphics[width=0.67\textwidth]{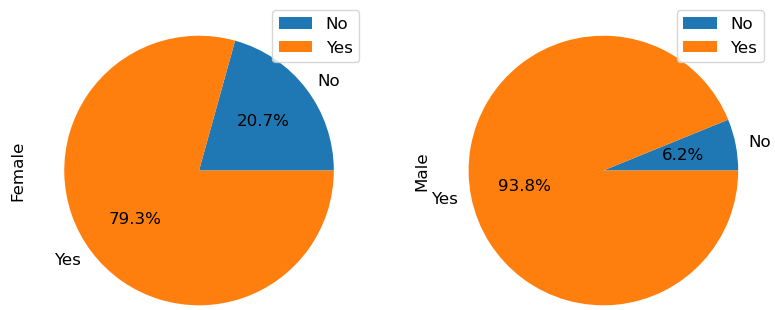}
    \caption{ChatGPT use by gender based on answers to question item number 3.}
    \label{fig:gender_use0}
\end{figure}

As shown in Figure \ref{fig:gender_use0}, while the proportion of LLM users exceeds 75\% for both genders, male students are significantly more likely to use these tools than female students (93.8\% vs 79.3\%, $\chi^2=16.84$, p-value $<10^{-4}$).

\begin{figure}[!h]
    \centering
    \includegraphics[width=0.67\textwidth]{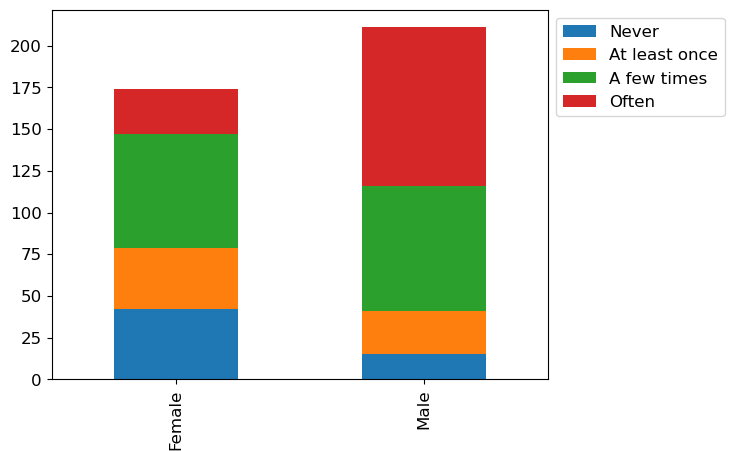}
    \caption{ChatGPT use by gender based on answers to question items number 4.1 and 5.1.}
    \label{fig:gender_use1}
\end{figure}

Furthermore, when we examine the frequency with which ChatGPT users of different genders utilise this LLM for academic purposes, the results, presented in Figure \ref{fig:gender_use1} and Table \ref{tab:gender_fu}, confirm that male and female students use LLMs at significantly different frequencies ($\chi^2=49.86$, p-value $<10^{-10}$). Notably, female students who use ChatGPT frequently are much fewer than their male counterparts. Specifically, 35\% of the chi-squared contribution comes from negative residuals for female students, compared to 23\% for male students, with positive residuals (See Table \ref{tab:gender_fur}). Additionally, when examining the chi-squared residuals and contributions for infrequent usage of LLMs, we observe that there are far fewer male students who use the tool infrequently compared to female students.

\begin{table}[!h]
\centering
\begin{tabular}{lrr}
 \toprule
 & Female & Male \\
\midrule
At least once & 37 & 26  \\
A few times & 68 & 75  \\
Often & 27 & 95  \\
\midrule
Never & 36 (+5) & 13 (+1) \\
\bottomrule
\end{tabular}    
\caption{Gender repartition and frequency of use (N=385).}
\label{tab:gender_fu}
\end{table}

\begin{table}[!h]
\centering
\begin{tabular}{lrr}
 \toprule
 & Female & Male \\
\midrule
At least once & 2.3 (19\%) & -1.9 (13\%) \\
A few times & 1.4 (7\%) & -1.1 (4\%)  \\
Often & -3.2 (35\%) & 2.6 (23\%)  \\
\bottomrule
\end{tabular}    
\caption{Residuals and chi squared contribution (between parenthesis) for the gender repartition and frequency of use, excluding non-users (N=328).}
\label{tab:gender_fur}
\end{table}

\begin{figure}[!h]
    \centering
    \includegraphics[width=0.67\textwidth]{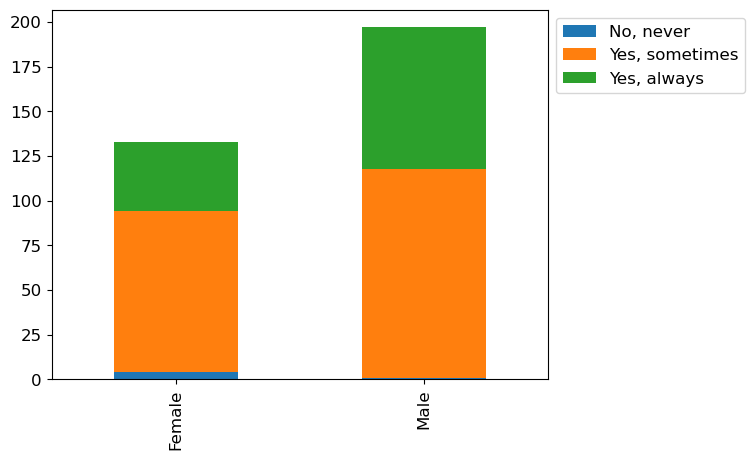}
    \caption{Reworking LLMs answers based on gender using question items 4.3 and 5.3, excluding non-users.}
    \label{fig:gender_rework}
\end{figure}

Following a similar approach to the previous sub-study, we now examine the frequency with which each gender revises answers generated by LLMs, such as ChatGPT. The results, presented in Figure \ref{fig:gender_rework} and based on Table \ref{tab:gender_rw}, indicate that male students are more likely than their female counterparts to systematically revise LLM-generated responses (40.1\% vs 29.3\%, respectively). With $\chi^2=6.72$ and a p-value of 0.035, we can confidently state that male and female students exhibit distinct habits in reworking and proofreading content generated by LLMs. However, it is noteworthy that the 95\% confidence intervals for these habits, calculated using the binomial-normal law approximation, show some overlap. This suggests that while the differences are statistically significant, they are not particularly pronounced.

\begin{table}[!h]
\centering
\begin{tabular}{lrr}
 \toprule
 & Female & Male \\
\midrule
No Never & 4 & 1  \\
Yes, sometimes & 90 & 117  \\
Yes, always & 39 & 79  \\
\midrule
Non-Users & 36 (+5) & 13 (+1) \\
\bottomrule
\end{tabular}    
\caption{Gender repartition and reworking the answers, excluding non-users (N=385).}
\label{tab:gender_rw}
\end{table}

\begin{figure}[!h]
    \centering
    \includegraphics[width=0.67\textwidth]{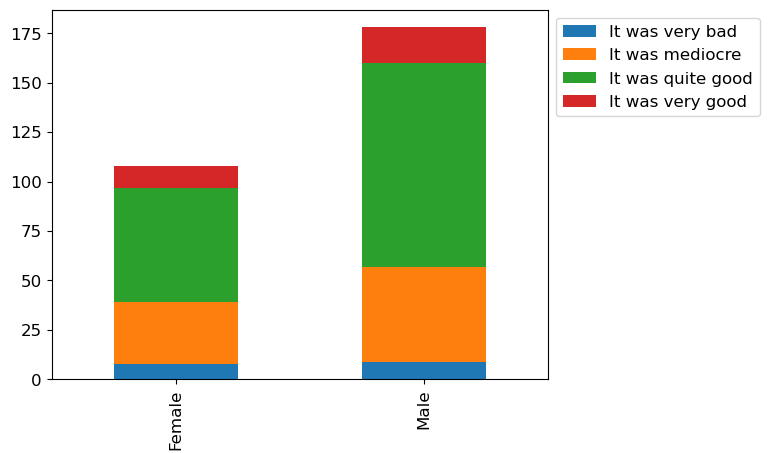}
    \caption{Answers average satisfaction based on the gender using question items 4.2 and 5.2, excluding non-users.}
    \label{fig:gender_sat}
\end{figure}

Then, Figure \ref{fig:gender_sat} and Table \ref{tab:gender_sat} show that satisfaction with LLM-generated responses is consistent between genders ($\chi^2=0.91$, p-value = 0.82), and both groups demonstrate similar satisfaction trends to those highlighted in the previous sub-study.

\begin{table}[!h]
\centering
\begin{tabular}{lrr}
 \toprule
 & Female & Male \\
\midrule
It was very bad & $7.4\%$ & $5.1\%$  \\
It was mediocre & $28.7\%$ & $27\%$  \\
It was quite good & $53.7\%$ & $57.9\%$  \\
It was quite bad & $10.2\%$ & $10.1\%$  \\
\bottomrule
\end{tabular}    
\caption{ChatGPT answer average satisfaction by gender, excluding non-users (N=328).}
\label{tab:gender_sat}
\end{table}

\begin{figure}[!h]
    \centering
    \includegraphics[width=0.67\textwidth]{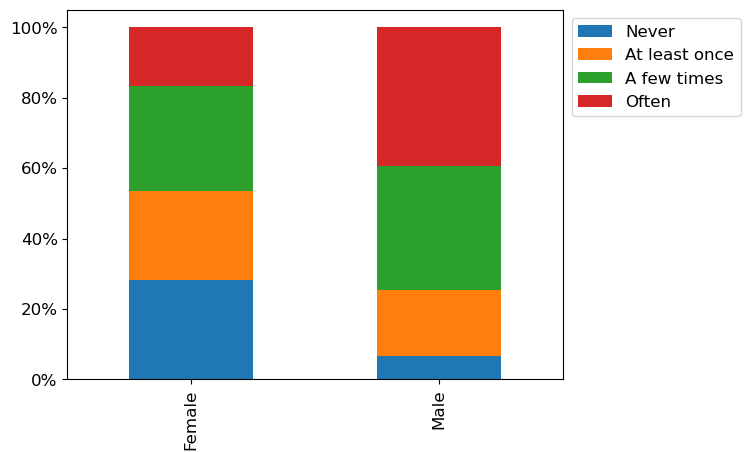}
    \caption{Use of ChatGPT for scientific related topics depending on the gender.}
    \label{fig:gender_ts}
\end{figure}

\begin{figure}[!h]
    \centering
    \includegraphics[width=0.67\textwidth]{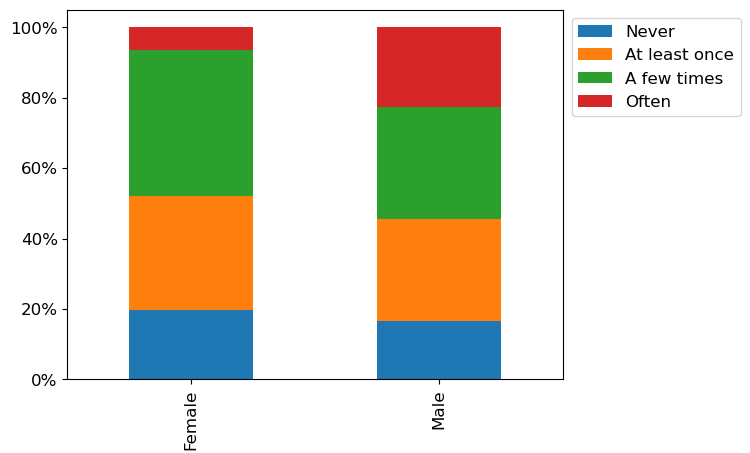}
    \caption{Use of ChatGPT for Humanities related topics depending on the gender.}
    \label{fig:gender_th}
\end{figure}

\begin{figure}[!h]
    \centering
    \includegraphics[width=0.75\textwidth]{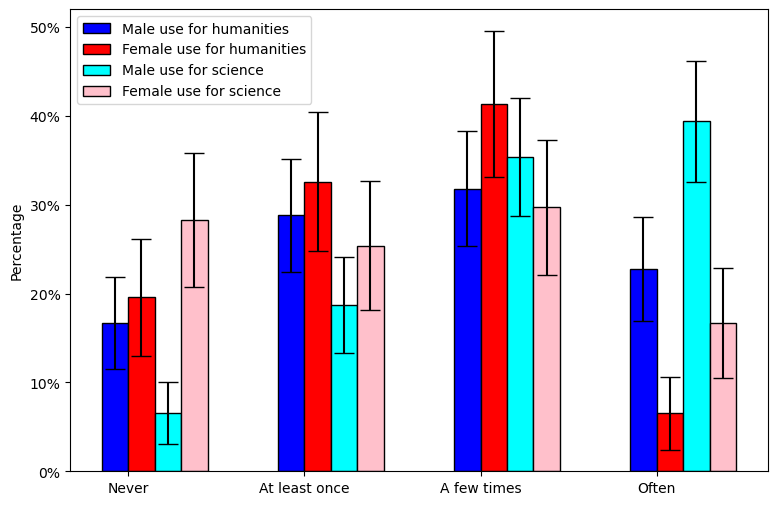}
    \caption{LLMs frequency of use by gender between science and humanities topic, with 95\% confidence intervals.}
    \label{fig:gender_hs_global}
\end{figure}

Finally, to ensure completeness and further investigate potential gender differences in the use of LLMs across various topics, we analysed responses to questionnaire items 2, 3, 4.1, and 5.1. Specifically, we examined whether there were significant differences in the topics for which students utilised ChatGPT (or other LLMs) based on their gender. The results are summarised in Table \ref{tab:gender_hs_dif}, with corresponding visualizations in Figures \ref{fig:gender_ts}, \ref{fig:gender_th} and \ref{fig:gender_hs_global}.

\begin{table}[!h]
\centering
\begin{tabular}{lcccc}
\toprule
 & \multicolumn{2}{c}{\emph{Science}} & \multicolumn{2}{c}{\emph{Humanities}} \\    
 & Female & Male & Female & Male \\
\midrule
Never & 39 & 13 & 27 & 33 \\
At least Once & 35 & 37 & 45 & 57 \\
A few times & 41 & 70 & 57 & 63 \\
Often & 23 & 78 & 9 & 45 \\
\bottomrule
\end{tabular}    
\caption{Use of ChatGPT by topic and by gender (N=336).}
\label{tab:gender_hs_dif}
\end{table}

Chi-squared tests yielded $\chi^2=41.18$ and p-value$=5.98 \times 10^{-9}$ for science-related topics and $0.0011$ for humanities, clearly indicating significant gender-based disparities in the use of LLMs such as ChatGPT. The residual analysis of ChatGPT usage for science topics reveals a notable surplus of female students who never use the tool for such purposes, in stark contrast to an excess of male students reporting frequent use. 

Furthermore, combined together the ``Often'' and ``Never'' answers in Sciences make for $94\%$ of the chi-squared distribution. Yet, the residual analysis for humanities still shows a deficit of female students that often use ChatGPT and the combined ``Often'' answers for males and females make for $83\%$ of the chi-squared contribution. All results are detailed in Table \ref{tab:gender_hs_res}. 
Lastly, when looking at Figure \ref{fig:gender_hs_global}, we can see that several confidence intervals are disjointed for male and female users in the ``Never'' and ``Often'' categories, which reinforces the previous findings.

While external factors, such as the higher tendency of male students to pursue science-oriented studies compared to their female counterparts \cite{wang2017gender}, undoubtedly contribute to this disparity, the emerging trend of women being less likely to use AI tools —particularly for science-related tasks— is concerning. This trend poses a potential challenge for the future representation of women in science and technology fields.

\begin{table}[!h]
\centering
\begin{tabular}{lcccc}
\toprule
 & \multicolumn{2}{c}{\emph{Science}} & \multicolumn{2}{c}{\emph{Humanities}} \\    
 & Female & Male & Female & Male \\
\midrule
Never & 3.8 (35\%) & -3.2 (25\%) & 0.47 (1\%) & -0.4 (1\%) \\
At least Once & 1 (2\%) & -0.83 (2\%) & 0.48 (1\%) & -0.4 (1\%) \\
A few times & -0.68 (1\%) & 0.57 (1\%) & 1.1 (7\%) & -0.92 (5\%) \\
Often & -2.9 (20\%) & -2.4 (14\%) & -2.8 (49\%) & 2.3 (34\%) \\
\bottomrule
\end{tabular}    
\caption{Use of ChatGPT by topic and by gender: residuals and Chi-squared contribution  (N=336).}
\label{tab:gender_hs_res}
\end{table}

In conclusion, this gender-based sub-study reveals an important risk of an emerging \emph{AI gap} between male and female students. First, male students are noticeably more likely to use these tools and exhibit a greater tendency to systematically revise the generated responses, suggesting a higher level of familiarity with LLMs. Second, we can see significant differences in the topics males and females are using LLMs for, with males more like likely to use them for scientific topics, and females for humanities related topics. This means that LLMs are likely to widen not one, but two gender gaps. Nevertheless, satisfaction levels remain nearly identical across genders.

Technological disparities between males and females are not a new phenomenon. However, given their role in perpetuating workplace inequalities and the profound impact LLMs are already having on the job market, it is imperative to take proactive measures to ensure that both boys and girls develop an equal familiarity with these tools and their limitations from an early age. Addressing this issue is crucial to preventing a widening gender gap in AI-related skills and opportunities.

However this result also led us to consider another hypothesis, that these gender disparities could be related to a different ethical approach: females could be more concerned about academic integrity. This consideration found a positive feedback in the sub-survey we will analyze in \ref{subsec:nonusers}. 

\subsection{Science topics vs Humanity topics analysis}

In this subsection, we discuss the results of our survey regarding potential differences in the use of ChatGPT in humanities-related fields compared to science-related fields.

\begin{figure}[!h]
    \centering
    \includegraphics[width=0.67\textwidth]{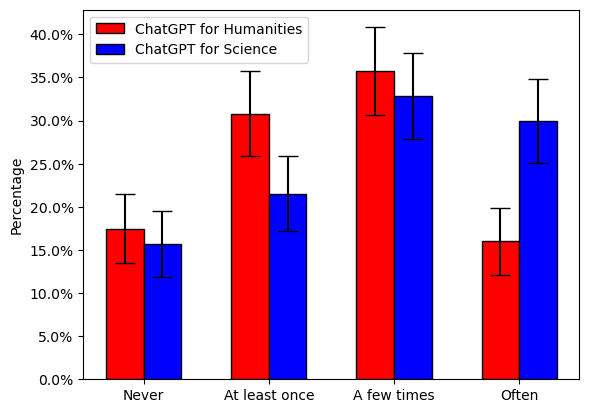}
    \caption{Comparing the use frequency of ChatGPT for Humanities and Science topics using question items 4.1 and 5.1, excluding non-users. Confidence intervals at 95\% are shown.}
    \label{fig:hs_use}
\end{figure}

\begin{figure}[!h]
    \centering
    \includegraphics[width=0.66\textwidth]{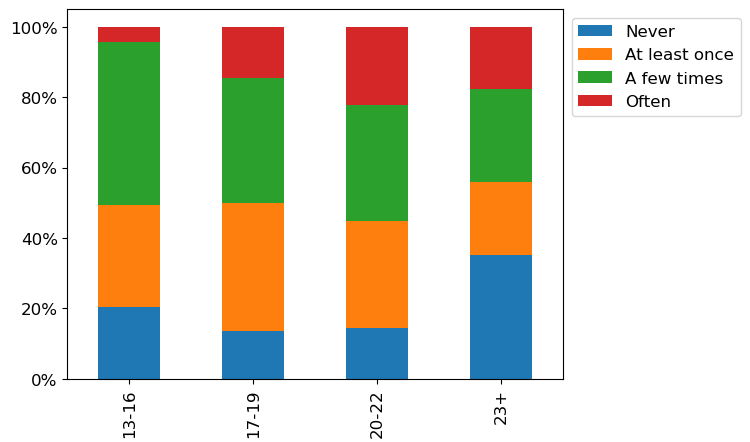}
    \caption{Raw usage frequencies for humanities topics based on item 4.1, excluding non-users.}
    \label{fig:hs_a1}
\end{figure}

\begin{figure}[!h]
    \centering
    \includegraphics[width=0.66\textwidth]{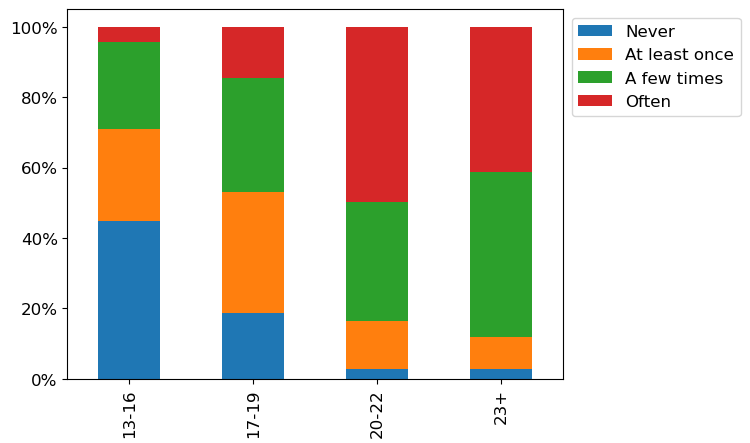}
    \caption{Raw usage frequencies for Science topics based on item 5.1, excluding non-users.}
    \label{fig:hs_a2}
\end{figure}

Using survey questions 4.1 and 5.1, we first assessed differences in the frequency of use between these two domains. The results are illustrated in Figure \ref{fig:hs_use} and summarised in Table \ref{tab:hs_use}. It is worth noting that Table \ref{tab:hs_use} columns derive from two different questions, not two modalities of the same question. Consequently, a chi-squared analysis was not deemed appropriate for this sub-study. Furthermore, as discussed in section \ref{sec:material}, our sample is biased due to a higher proportion of science students, a disparity evident in Figures \ref{fig:hs_a1} and \ref{fig:hs_a2}.

\begin{table}[!h]
\centering
\begin{tabular}{lccc}
 \toprule
 Answer & Numbers & Percentages & Percentages IC95 \\
\midrule
\midrule
\multicolumn{4}{c}{\emph{Humanities}} \\
\midrule
Never & 60 & $17.4\%$ & $[13.4\%;21.5\%]$ \\
At least once & 106 & $30.8\%$ & $[25.9\%;35.7\%]$ \\
A few times & 123 & $35.8\%$ & $[30.7\%;40.8\%]$ \\
Often & 55 & $16\%$ &  $[12.1\%;19.9\%]$ \\
\midrule
\multicolumn{4}{c}{\emph{Science}} \\
\midrule
Never & 54 & $15.7\%$ & $[11.9\%;19.5\%]$ \\
At least once & 74 & $21.5\%$ & $[17.2\%;25.9\%]$ \\
A few times & 113 & $32.8\%$ & $[27.9\%;37.8\%]$ \\
Often & 103 & $29.9\%$ & $[25.1\%;37.8\%]$ \\
\bottomrule
\end{tabular}    
\caption{Comparing the use frequency of ChatGPT for Humanities and Science topics, excluding non-users (N=344).}
\label{tab:hs_use}
\end{table}

To address these limitations, we used confidence intervals to assess differences in the use of ChatGPT between science and humanities fields. These intervals were computed using a normal approximation of the binomial distribution at a 95\% confidence level.

As can be seen from Figure \ref{fig:hs_use}, the results reveal overlapping confidence intervals for students who have never used ChatGPT or have used it only a few times in both fields. However, the intervals for students who have used it ``At least once'' or ``Often'' are disjoint, with humanities leading in the former category and sciences in the latter. Given that students who never use ChatGPT were excluded from this analysis, we infer that non-users in humanities and sciences are distinct populations. The disjoint intervals suggest a polarisation: humanities students are more likely to be occasional users, whereas science students dominate the frequent user category. This trend is likely influenced by the over representation of science students in our sample. Additionally, previous sub-studies have shown that secondary and high school students tend to use LLMs less. Thus, a broader interpretation might be that as students advance in their academic careers, they increasingly use LLMs in their specialised fields while their use in other domains becomes more occasional.

\begin{figure}[!h]
    \centering
    \includegraphics[width=0.67\textwidth]{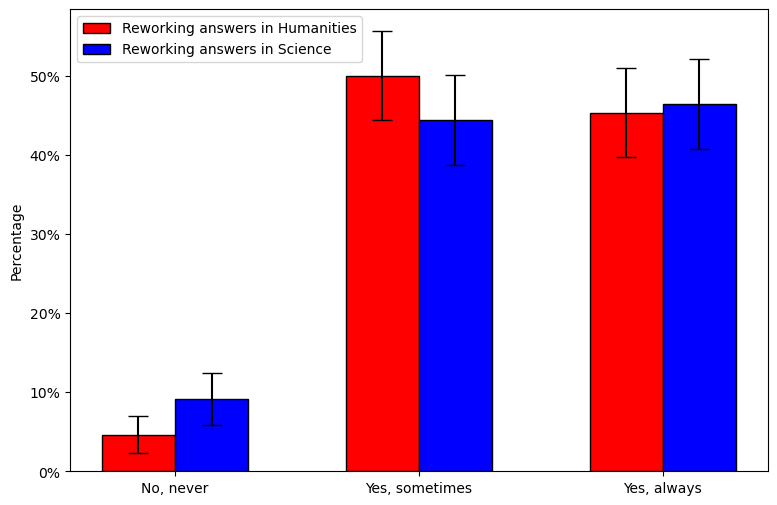}
    \caption{Percentages of students using LLMs that rework answers in humanities and science based on question items 4.3 and 5.3, excluding non-users. Confidence intervals at 95\% are shown.}
    \label{fig:hs_rw}
\end{figure}

In Figure \ref{fig:hs_rw} and Table \ref{tab:hs_rw}, we examine whether students rework LLM-generated answers differently when addressing humanities versus science topics. The results align with previous sub-studies: fewer than 10\% of students report not reworking LLM-generated answers at all, while slightly less than 50\% systematically rework them. Additionally, the confidence intervals for reworking habits between the two fields overlap entirely, suggesting no significant difference in this behaviour between humanities and sciences. The bias towards more science students in the sample does not appear to have impacted these findings, indicating that good reworking habits are consistent regardless of the field of study.

\begin{table}[!h]
\begin{tabular}{lcc}
 \toprule
 & \emph{Humanities} & \emph{Sciences} \\
\midrule
No, never & 14 & 27 \\
Yes, sometimes & 151 & 131  \\
Yes, always & 137 & 137  \\
\midrule
Did not use it & 42 & 49 \\
\bottomrule
\end{tabular}   
\caption{Raw numbers of students reworking LLMs answers in Humanities and Science (N=344).}
\label{tab:hs_rw}
\end{table}

\begin{figure}[!h]
    \centering
    \includegraphics[width=0.67\textwidth]{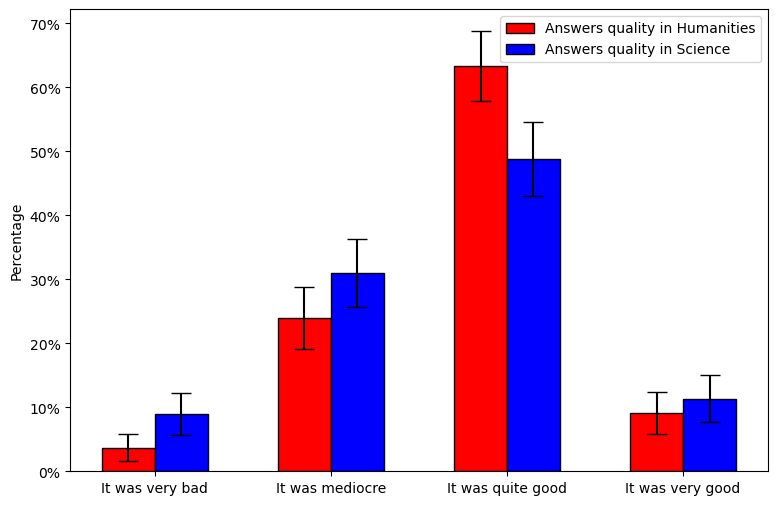}
    \caption{Perception of LLM answers quality in humanities versus science topics based on question items 4.2 and 5.2, excluding non-users. Confidence intervals at 95\% are shown.}
    \label{fig:hs_sat}
\end{figure}

Finally, Figure \ref{fig:hs_sat} and Table \ref{tab:hs_sat} show how satisfied students were with LLM-generated answers based on the type of topic. The confidence intervals for ``It was very bad'' and ``It was quite good'' are disjoint, while those for ``It was mediocre'' and ``It was very good'' overlap. Interestingly, satisfaction appears higher for humanities topics on average, with the exception of the ``It was very good'' category, where science topics slightly lead.

In our previous sub-studies, we observed that overall satisfaction with LLM-generated answers exceeded 50\%. This sub-study confirms that this trend holds true across both humanities and sciences. While the chi-squared test (p-value $\approx$ 0.0016) indicates independent behaviours between the two fields, the lack of clear trends in residuals and confidence intervals, combined with the known bias of a greater number of science students, limits further interpretation.

\begin{table}[!h]
\begin{tabular}{lcccc}
 \toprule
 Answer & Numbers & $\chi^2$ norm. residuals & Percentages & Percentages IC95 \\
\midrule
\midrule
\multicolumn{5}{c}{\emph{Humanities}} \\
\midrule
It was very bad & 11 & -1.8 & $3.7\%$ & $[1.6\%;5.9\%]$ \\
It was mediocre & 71 & -1.1 & $23.9\%$ & $[19.1\%;28.8\%]$ \\
It was quite good & 188 & 1.7 & $63.3\%$ & $[57.8\%;68.8\%]$ \\
It was very good & 27 & -0.6 & $9.1\%$ &  $[5.8\%;12.3\%]$ \\
\midrule
\multicolumn{5}{c}{\emph{Science}} \\
\midrule
It was very bad & 26 & 1.8 & $8.9\%$ & $[5.7\%;12.2\%]$ \\
It was mediocre & 90 & 1.2 & $30.9\%$ & $[25.6\%;36.2\%]$ \\
It was quite good & 142 & -1.7 & $48.8\%$ & $[43.1\%;54.5\%]$ \\
It was very good & 33 & 0.61 & $11.3\%$ &  $[7.7\%;15\%]$ \\
\bottomrule
\end{tabular}    
\caption{Comparing ChatGPT students' satisfaction for Humanities and Science topics, excluding non-users (N=344).}
\label{tab:hs_sat}
\end{table}

From this third sub-study, we draw the following conclusions:
\begin{itemize}
    \item the habit of reworking LLM-generated answers is independent of the topic type and remains consistently high;
    \item as students advance in their studies, their use of ChatGPT becomes more frequent in their primary fields and more occasional in other areas;
    \item while differences exist in the perceived quality of LLM-generated answers depending on the topic, no clear trends or explanations can be derived solely from this analysis. 
\end{itemize}

\subsection{Analysis of device preferences}

\begin{figure}[!h]
    \centering
    \includegraphics[width=0.67\textwidth]{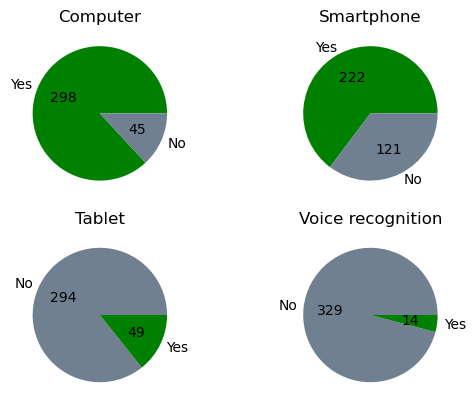}
    \caption{Devices used to prompt ChatGPT based on question item 6.}
    \label{fig:devices}
\end{figure}

Using question item 6, we established Figure \ref{fig:devices} and Table \ref{tab:devices}. We remind that question 6 was multiple choice with no obligation to answer. As such, some user checked several devices, and only 343 of the 346 self-reported users chose to answer this item.

\begin{table}[!h]
\begin{tabular}{lcccc}
 \toprule
 & Computer & Smartphone & Tablet & Voice recognition \\
\midrule
Yes & 298 & 222 & 49 & 14 \\
No & 45 & 121 & 294 & 329 \\
\bottomrule
\end{tabular}    
\caption{Report of devices used to prompt ChatGPT (N=343).}
\label{tab:devices}
\end{table}

First, we can see that the use of voice recognition is not yet democratized, with only around 4\% of our respondents that used it. Then, we can see that both computers (or laptops) and smartphones are frequently used by students to prompt ChatGPT, with a preference for computers (87\% vs 65\%). However, as we can see from Tables \ref{tab:devicesC} and \ref{tab:devicesSP}, the situation is not homogeneous between genders and age categories.

\begin{table}[!h]
\begin{tabular}{lcccccc}
 \toprule
 & Female & Male & 13-16 & 17-19 & 20-22 & 23+ \\
\midrule
Yes & 104 & 188 & 41 & 82 & 142 & 33 \\
No & 33 & 10 & 27 & 14 & 3 & 1 \\
\bottomrule
\end{tabular}    
\caption{Computer to prompt ChatGPT depending on age and gender (N=343).}
\label{tab:devicesC}
\end{table}

Using chi-squared analysis, we can see that only stable trend is the use of smartphone accross genders ($\chi^2=1.22$, p-value=$0.27$). On the other hand, the use of smartphone is tied to the age category ($\chi^2=60.97$, p-value$<10^{-12}$) with the two youngest categories being the most frequent users (the 13-16 and 17-19yo age range). Likewise, the use of computers and laptops depends on both the age ($\chi^2=17.29$, p-value=$6 \times 10^{-4}$) and the gender ($\chi^2=24.55$, p-value=$7.2 \times 10^{-7}$), with older students (20-22 and 23+) as well as male students being significantly more likely to use a computer than younger students and female students.

\begin{table}[!h]
\begin{tabular}{lcccccc}
 \toprule
 & Female & Male & 13-16 & 17-19 & 20-22 & 23+ \\
\midrule
Yes & 94 & 123 & 56 & 67 & 80 & 19 \\
No & 43 & 75 & 12 & 29 & 65 & 15 \\
\bottomrule
\end{tabular}    
\caption{Smartphone to prompt ChatGPT depending on age and gender. (N=343)}
\label{tab:devicesSP}
\end{table}

These results are coherent with known trends where the younger generation is more likely to use portable devices than personal computers. For now the computer remains dominant as a tool to use ChatGPT because university students use it more often than high school students, but we can clearly see that this should change in the next 4 years when the younger generation enters higher education. This point must be taken strongly into account to re-think educational tools and homeworks in next future.

\subsection{Analysis of non-chatGPT users}\label{subsec:nonusers}

In this subsection, we analyse the responses of the 51 participants who reported never having used ChatGPT in an academic context, focusing on their answers to question item 9. The results are presented in Table \ref{tab:non_users}. It is important to note that item 9 allowed multiple answers, which explains why the total exceeds 51, and also included a free-text option. In this survey, two respondents indicated they were unaware that ChatGPT offered a freemium model.

\begin{table}[!h]
\begin{tabular}{lrr}
 \toprule
Reason given & Numbers & Frequencies \\
\midrule
I did not think it was useful & 27 & $52.9\%$ \\
I do not think it is honest in academia & 22 & $43.13\%$ \\
I did not know this tool & 5 & $9.8\%$ \\
I did not think about using it for academia & 8 & $15.7\%$ \\
Other: I did not know it was free & 2 & $3.9\%$ \\
\bottomrule
\end{tabular}    
\caption{Reasons given by students that declared that they never used ChatGPT or any other LLM. (N=51)}
\label{tab:non_users}
\end{table}

\begin{figure}[!h]
    \centering
    \includegraphics[width=0.67\textwidth]{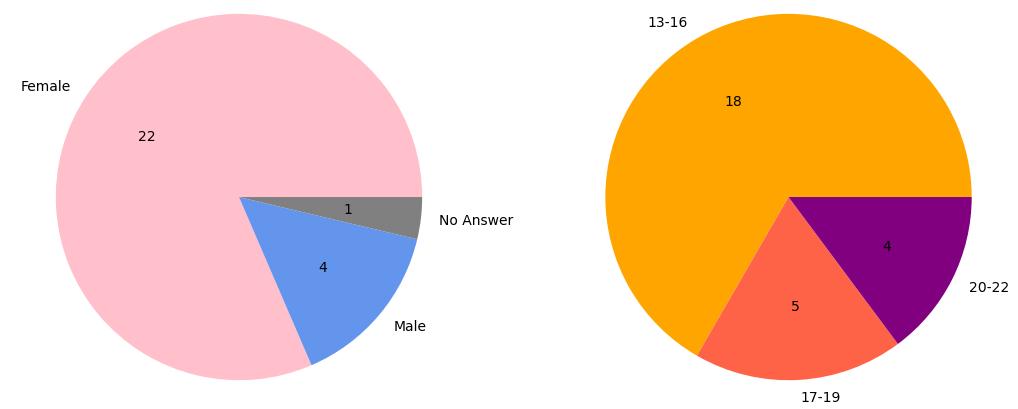}
    \caption{Share of students that said they do not use ChatGPT because they don't find it useful.}
    \label{fig:not_useful}
\end{figure}

\begin{figure}[!h]
    \centering
    \includegraphics[width=0.67\textwidth]{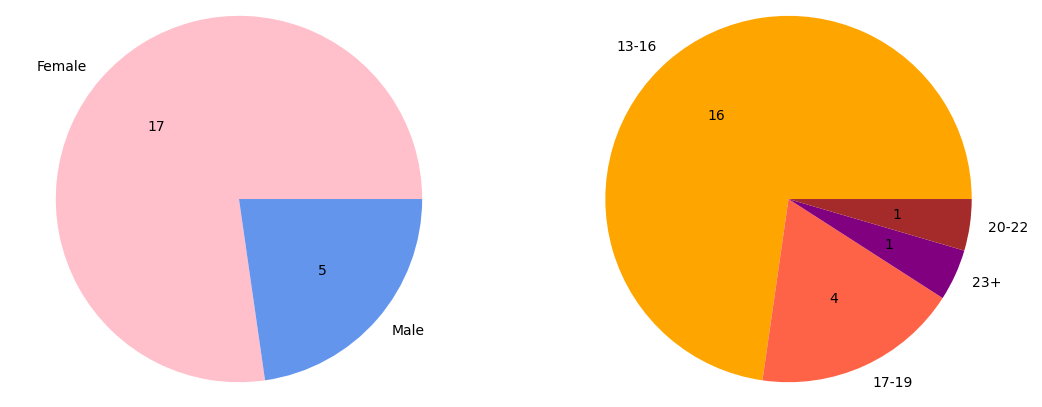}
    \caption{Share of students that said they do not use ChatGPT in academia because they find it dishonest to do so.}
    \label{fig:dishonest}
\end{figure}

While the sample size is small, we did not observe any notable trends among respondents who indicated that they ``did not know [ChatGPT]'', ``did not think about using it in academia'', or mistakenly believed it was not free. However, clear patterns emerged in the responses to the statements ``I do not think [ChatGPT] is useful'' and ``I do not think it is honest in academia'', as shown in Figures \ref{fig:not_useful} and \ref{fig:dishonest}.

In both cases, young women aged 13 to 16 years comprised the majority of respondents (over 66\%), which is not a surprise given the results of previous sub-studies and proportions shown in Figures \ref{fig:age_use0}, \ref{fig:age_rework}, \ref{fig:gender_use0} and \ref{fig:gender_rework}. On one hand, this suggests an encouraging level of ethical reflection, with young women questioning the morality of using such tools in academia: this confirms what we hypothesized in section \ref{sec:gender} that is that male and female students could have different ethical points of view when it comes to using LLMs in academia, thus leading to diverging frequencies of use. This point must be further investigated in future works to better understand if this disparity is mostly related to ethical issues of if it highlights -again- a possible risk of a gender gap in AI literacy and the ability to effectively utilise LLMs.

\subsection{Comments on non-academic ChatGPT uses by student populations}

In question item 8, we asked respondents whether they used ChatGPT or other LLMs for non-academic purposes. This was a multiple-choice question that included several pre-written options, such as ``for personal documents and text writing", ``to write administrative documents (e.g., CVs, motivation letters)", and ``for social networks". Additionally, an open text field allowed respondents to specify other uses. The results are presented in Table \ref{tab:others}, where the line ``As a search engine'' includes all uses of ChatGPT related to online research, cultural queries, or daily questions that could otherwise have been addressed through traditional search engines or platforms like Wikipedia.

\begin{table}[!h]
\begin{tabular}{lrr}
 \toprule
Other uses & Numbers & Percentages \\
\midrule
Personal documents and texts writing & 163 & $41.2\%$ \\
Administrative document writing & 131 & $33.2\%$ \\
For social networks & 74 & $18.7\%$ \\
As a search engine & 67 & $17\%$ \\
Other uses & 8 & $2\%$ \\
\bottomrule
\end{tabular}    
\caption{Non-academic uses of ChatGPT. (N=395)}
\label{tab:others}
\end{table}

With the unsurprising exception of ``administrative document writing'', which was less commonly reported by younger respondents, no significant (or new) age or gender trends were observed for this question item.

The most frequent non-academic use of ChatGPT among respondents was personal document and text writing, with 41.2\% of students indicating that they had used LLMs for this purpose. However, overall percentages for non-academic use were surprisingly lower than anticipated. One possible explanation is that academic tasks, such as homework and research for school, are the primary use cases for ChatGPT, as nearly 60\% of respondents reported not using it for anything else. Alternatively, the omission of ``as a search engine" as a pre-written option in the multiple-choice answers may have inadvertently introduced bias, overlooking what could be another major use of LLMs.

In any case, we recommend caution when interpreting the results of this question item, as it could warrant further investigation in future studies.

\section{Conclusions and Future works}
\label{sec:conclusions}

In this paper, we have provided a comprehensive analysis of how students have integrated on their own the use of LLMs such as ChatGPT into their academic routines. This analysis is based on a survey conducted in France and Italy among students aged 13 to 25 years old. 
The survey had three main objectives:

\begin{itemize}
    \item Assessing the pervasiveness of LLMs across different student populations based on age and gender, with the aim of evaluating how systemic these tools are in academic life.
    \item Analysing potential differences in usage depending on students' age, gender, and the type of academic topic. By \emph{differences in usage}, we refer to factors such as frequency of use, perceived trustworthiness, proofreading and reformulation habits, as well as device preferences.
    \item Identifying opportunities, threats, and risks related to learning challenges and technological gaps between LLM users and non-users.
\end{itemize}

Our findings revealed the widespread use of LLMs across all age groups and topics, with notable trends such as the already frequent use by two-thirds of students aged 13 to 16, and an increase in usage frequency as students progress through their curriculum. 

However, significant gender differences have emerged, with male students showing higher usage rates in both humanities and scientific fields. The gap in usage frequencies is particularly pronounced for scientific applications of LLMs. While this may reflect a pre-existing and culturally exacerbated propensity of male students toward scientific topics, it is evident that LLMs could widen the technological gender gap. This raises urgent concerns about widening disparities in AI literacy and gender equality, which could extend beyond STEM and AI-intensive fields to affect nearly all employment sectors impacted by AI. Addressing these disparities should become a priority for educational policymakers.
Taking into account the results shown in sections \ref{sec:gender} and \ref{subsec:nonusers}, we found that these disparities may also be related to a different ethical approach: females could be more concerned about academic integrity. 
We believe that this issues would deserve deeper investigation to better comprehend whether this disparity is primarily linked to differing ethical attitudes or rather to unequal access, propensity or education regarding computing and AI resources.

The widespread popularity of these tools across all student populations highlights a shift in the educational paradigm, with traditional assessment methods such as homework increasingly subject to AI intervention. This calls for a reassessment of pedagogical strategies to incorporate AI tools constructively and ensure cognitive engagement rather than simple delegation to an AI. Our results also demonstrate the need to teach students—especially younger generations—to critically evaluate and revise outputs from these tools, as this habit is far from systematic at present.

Although it is beyond the scope of this paper, we must note the ethical challenges and risks associated with the over-reliance on AI tools, as well as their potential to erode critical thinking and foundational cognitive skills. Our findings indicate that older students are more likely to critically analyse AI-generated results, likely because they developed these skills during formative years when AI assistance was not yet available, but also because they saw the emergence and evolution of these tools and had time to learn about their strengths and weaknesses. Future students, exposed to AI from a younger age, may struggle to develop the same level of critical assessment and this must be taken into account. Because previous research have shown that typing is less effective than handwriting for words and concept retention \cite{naka1995effect,van2020importance}, it is very likely that copy-pasting LLM-generated answers -and reworking them a bit in best case scenarii- offers even less cognitive benefit. Furthermore, things might get worse if voice recognition takes off, which is not yet the case according to our results.

In conclusion, our study raises numerous questions and confirms several concerns that should be addressed by educational systems. On the one hand, AI and LLMs represent a tremendous opportunity, and their mastery will become indispensable in the job market. On the other hand, current educational structures appear ill-equipped to mitigate the potential harms of these tools, including gender gaps, AI illiteracy, and the erosion of critical thinking and foundational cognitive skills.

Given that no age group (not even secondary and high-schools) or subject area has remained untouched by LLMs, we strongly recommend that all educational institutions begin integrating these systems into their teaching methods. Key areas of focus should include ethics, critical thinking, the limitations of AI, and the environmental costs associated with these technologies. These topics can be introduced at an early age without overburdening already demanding curricula. Similarly, revising teaching and evaluation methods to account for LLMs is essential to safeguard learning outcomes.
By embracing these recommendations, educational systems can harness the potential of LLMs to enhance learning while mitigating risks, ensuring that these powerful tools serve as a bridge rather than a barrier to equitable and effective education.

Based on our survey results and the statements presented in this section, we have identified 2 distinct, and in some ways opposing, approaches to rethinking homework in light of the existence of LLMs:
\begin{itemize}
    \item The first approach would acknowledge the existence of LLMs and they likelihood that they will be used. It would consist in adding an unassisted restitution part to any homework, thus forcing students to appropriate the answers produced by these tools and in some case mobilize memorization skills.
    \item The second approach embraces the use of ChatGPT, particularly suited to humanities subjects. In this scenario, students would use ChatGPT to complete their assignments and subsequently evaluate their learning process. They would analyze the type of reasoning required, assess the elaboration needed for the task, and compare the time spent utilizing ChatGPT's responses versus solving the task independently.
\end{itemize}

These methodologies could be tested in a research setting, where classes from a specific discipline would engage with homework designed according to these approaches. Learning outcomes, as well as feedback from students and teachers, could then be assessed through pre- and post-tests to evaluate the effectiveness of each strategy.

Finally, while this falls outside the framework of our survey too, we believe that environmental issues related to ChatGPT use \cite{biswas2023potential,george2023environmental,zhu2023chatgpt} must be taken into account when considering adopting these tools as current part of an educational setting.

\backmatter

\bmhead{Acknowledgments}
We would like to thank all professors and heads of institutions that accepted to have their students participating in our survey. 
We also thank our colleague helped proof-reading the paper and making the tables more beautiful.

\section*{Declarations}

\subsection*{Fundings}
This study received no external funding.

\subsection*{Conflicts of interest}
The authors declare no conflict of interest.

\subsection*{Data and code availability}
All participants in the surveys presented in this paper were informed of their rights under the General Data Protection Regulation (GDPR - EU Regulation 2016/679 and Legislative Decree no. 196/2003) before taking part in the survey. The parents or legal representatives of underage participants received the same information and gave their informed consent for their child to participate.

In accordance with EU GDPR regulations, the data cannot be shared and will be erased at most at the end of the 2 years periods following the end of the survey.

The code could be made available upon request, but it presents little to no interest without the dataset.

\subsection*{Authors' contributions}
Both authors contributed equally.

\bibliography{biblio}


\begin{thebibliography}{68}
\ifx \bisbn   \undefined \def \bisbn  #1{ISBN #1}\fi
\ifx \binits  \undefined \def \binits#1{#1}\fi
\ifx \bauthor  \undefined \def \bauthor#1{#1}\fi
\ifx \batitle  \undefined \def \batitle#1{#1}\fi
\ifx \bjtitle  \undefined \def \bjtitle#1{#1}\fi
\ifx \bvolume  \undefined \def \bvolume#1{\textbf{#1}}\fi
\ifx \byear  \undefined \def \byear#1{#1}\fi
\ifx \bissue  \undefined \def \bissue#1{#1}\fi
\ifx \bfpage  \undefined \def \bfpage#1{#1}\fi
\ifx \blpage  \undefined \def \blpage #1{#1}\fi
\ifx \burl  \undefined \def \burl#1{\textsf{#1}}\fi
\ifx \doiurl  \undefined \def \doiurl#1{\url{https://doi.org/#1}}\fi
\ifx \betal  \undefined \def \betal{\textit{et al.}}\fi
\ifx \binstitute  \undefined \def \binstitute#1{#1}\fi
\ifx \binstitutionaled  \undefined \def \binstitutionaled#1{#1}\fi
\ifx \bctitle  \undefined \def \bctitle#1{#1}\fi
\ifx \beditor  \undefined \def \beditor#1{#1}\fi
\ifx \bpublisher  \undefined \def \bpublisher#1{#1}\fi
\ifx \bbtitle  \undefined \def \bbtitle#1{#1}\fi
\ifx \bedition  \undefined \def \bedition#1{#1}\fi
\ifx \bseriesno  \undefined \def \bseriesno#1{#1}\fi
\ifx \blocation  \undefined \def \blocation#1{#1}\fi
\ifx \bsertitle  \undefined \def \bsertitle#1{#1}\fi
\ifx \bsnm \undefined \def \bsnm#1{#1}\fi
\ifx \bsuffix \undefined \def \bsuffix#1{#1}\fi
\ifx \bparticle \undefined \def \bparticle#1{#1}\fi
\ifx \barticle \undefined \def \barticle#1{#1}\fi
\bibcommenthead
\ifx \bconfdate \undefined \def \bconfdate #1{#1}\fi
\ifx \botherref \undefined \def \botherref #1{#1}\fi
\ifx \url \undefined \def \url#1{\textsf{#1}}\fi
\ifx \bchapter \undefined \def \bchapter#1{#1}\fi
\ifx \bbook \undefined \def \bbook#1{#1}\fi
\ifx \bcomment \undefined \def \bcomment#1{#1}\fi
\ifx \oauthor \undefined \def \oauthor#1{#1}\fi
\ifx \citeauthoryear \undefined \def \citeauthoryear#1{#1}\fi
\ifx \endbibitem  \undefined \def \endbibitem {}\fi
\ifx \bconflocation  \undefined \def \bconflocation#1{#1}\fi
\ifx \arxivurl  \undefined \def \arxivurl#1{\textsf{#1}}\fi
\csname PreBibitemsHook\endcsname

\bibitem[\protect\citeauthoryear{Minaee et~al.}{2024}]{minaee2024}
\begin{botherref}
\oauthor{\bsnm{Minaee}, \binits{S.}},
\oauthor{\bsnm{Mikolov}, \binits{T.}},
\oauthor{\bsnm{Nikzad}, \binits{N.}},
\oauthor{\bsnm{Chenaghlu}, \binits{M.}},
\oauthor{\bsnm{Socher}, \binits{R.}},
\oauthor{\bsnm{Amatriain}, \binits{X.}},
\oauthor{\bsnm{Gao}, \binits{J.}}:
Large language models: A survey
(2024)
\end{botherref}
\endbibitem

\bibitem[\protect\citeauthoryear{OpenAI}{2022}]{openai2023chatgpt}
\begin{botherref}
\oauthor{\bsnm{OpenAI}}:
Introducing ChatGPT
(2022).
\url{https://openai.com/index/chatgpt/}
\end{botherref}
\endbibitem

\bibitem[\protect\citeauthoryear{Burton et~al.}{2024}]{burton2024large}
\begin{botherref}
\oauthor{\bsnm{Burton}, \binits{J.W.}},
\oauthor{\bsnm{Lopez-Lopez}, \binits{E.}},
\oauthor{\bsnm{Hechtlinger}, \binits{S.}},
\oauthor{\bsnm{Rahwan}, \binits{Z.}},
\oauthor{\bsnm{Aeschbach}, \binits{S.}},
\oauthor{\bsnm{Bakker}, \binits{M.A.}},
\oauthor{\bsnm{Becker}, \binits{J.A.}},
\oauthor{\bsnm{Berditchevskaia}, \binits{A.}},
\oauthor{\bsnm{Berger}, \binits{J.}},
\oauthor{\bsnm{Brinkmann}, \binits{L.}}, et al.:
How large language models can reshape collective intelligence.
Nature Publishing Group UK London
(2024)
\end{botherref}
\endbibitem

\bibitem[\protect\citeauthoryear{Luo et~al.}{2024}]{luo2024large}
\begin{botherref}
\oauthor{\bsnm{Luo}, \binits{X.}},
\oauthor{\bsnm{Rechardt}, \binits{A.}},
\oauthor{\bsnm{Sun}, \binits{G.}},
\oauthor{\bsnm{Nejad}, \binits{K.K.}},
\oauthor{\bsnm{Y{\'a}{\~n}ez}, \binits{F.}},
\oauthor{\bsnm{Yilmaz}, \binits{B.}},
\oauthor{\bsnm{Lee}, \binits{K.}},
\oauthor{\bsnm{Cohen}, \binits{A.O.}},
\oauthor{\bsnm{Borghesani}, \binits{V.}},
\oauthor{\bsnm{Pashkov}, \binits{A.}}, et al.:
Large language models surpass human experts in predicting neuroscience results
(2024)
\end{botherref}
\endbibitem

\bibitem[\protect\citeauthoryear{Toner-Rodgers}{2024}]{TonerRodgers2024}
\begin{bchapter}
\bauthor{\bsnm{Toner-Rodgers}, \binits{A.}}:
\bctitle{{Artificial Intelligence, Scientific Discovery, and Product Innovation}}.
In: \bbtitle{National Bureau of Economic Research}
(\byear{2024}).
\burl{https://conference.nber.org/conf_papers/f210475.pdf}
\end{bchapter}
\endbibitem

\bibitem[\protect\citeauthoryear{Baskara et~al.}{2023}]{baskara2023exploring}
\begin{barticle}
\bauthor{\bsnm{Baskara}, \binits{R.}}, \betal:
\batitle{Exploring the implications of chatgpt for language learning in higher education.}
\bjtitle{Indonesian Journal of English Language Teaching and Applied Linguistics}
\bvolume{7}(\bissue{2}),
\bfpage{343}--\blpage{358}
(\byear{2023})
\end{barticle}
\endbibitem

\bibitem[\protect\citeauthoryear{Su et~al.}{2023}]{su2023collaborating}
\begin{barticle}
\bauthor{\bsnm{Su}, \binits{Y.}},
\bauthor{\bsnm{Lin}, \binits{Y.}},
\bauthor{\bsnm{Lai}, \binits{C.}}:
\batitle{Collaborating with chatgpt in argumentative writing classrooms}.
\bjtitle{Assessing Writing}
\bvolume{57},
\bfpage{100752}
(\byear{2023})
\end{barticle}
\endbibitem

\bibitem[\protect\citeauthoryear{Lee and Yeo}{2022}]{lee2022developing}
\begin{barticle}
\bauthor{\bsnm{Lee}, \binits{D.}},
\bauthor{\bsnm{Yeo}, \binits{S.}}:
\batitle{Developing an ai-based chatbot for practicing responsive teaching in mathematics}.
\bjtitle{Computers \& Education}
\bvolume{191},
\bfpage{104646}
(\byear{2022})
\end{barticle}
\endbibitem

\bibitem[\protect\citeauthoryear{Ayala}{2023}]{ayala2023chatgpt}
\begin{barticle}
\bauthor{\bsnm{Ayala}, \binits{S.}}:
\batitle{Chatgpt as a universal design for learning tool supporting college students with disabilities.}
\bjtitle{Educational Renaissance}
\bvolume{12},
\bfpage{22}--\blpage{41}
(\byear{2023})
\end{barticle}
\endbibitem

\bibitem[\protect\citeauthoryear{Tamdjidi et~al.}{2023}]{tamdjidi2023chatgpt}
\begin{botherref}
\oauthor{\bsnm{Tamdjidi}, \binits{R.}}, et al.:
ChatGPT as an assistive technology to enhance reading comprehension for individuals with ADHD
(2023)
\end{botherref}
\endbibitem

\bibitem[\protect\citeauthoryear{Rahman and Watanobe}{2023}]{rahman2023chatgpt}
\begin{barticle}
\bauthor{\bsnm{Rahman}, \binits{M.M.}},
\bauthor{\bsnm{Watanobe}, \binits{Y.}}:
\batitle{Chatgpt for education and research: Opportunities, threats, and strategies}.
\bjtitle{Applied Sciences}
\bvolume{13}(\bissue{9}),
\bfpage{5783}
(\byear{2023})
\end{barticle}
\endbibitem

\bibitem[\protect\citeauthoryear{Bentley et~al.}{2024}]{bentley2024digital}
\begin{botherref}
\oauthor{\bsnm{Bentley}, \binits{S.V.}},
\oauthor{\bsnm{Naughtin}, \binits{C.K.}},
\oauthor{\bsnm{McGrath}, \binits{M.J.}},
\oauthor{\bsnm{Irons}, \binits{J.L.}},
\oauthor{\bsnm{Cooper}, \binits{P.S.}}:
The digital divide in action: how experiences of digital technology shape future relationships with artificial intelligence.
Springer
(2024)
\end{botherref}
\endbibitem

\bibitem[\protect\citeauthoryear{Herbert}{2017}]{herbert2017digital}
\begin{botherref}
\oauthor{\bsnm{Herbert}, \binits{S.}}:
Digital development and the digital gender gap
(2017)
\end{botherref}
\endbibitem

\bibitem[\protect\citeauthoryear{Longoria et~al.}{2022}]{longoria2022systematic}
\begin{barticle}
\bauthor{\bsnm{Longoria}, \binits{I.A.-I.}},
\bauthor{\bsnm{Bustamante-Bello}, \binits{R.}},
\bauthor{\bsnm{Ram{\'\i}rez-Montoya}, \binits{M.S.}},
\bauthor{\bsnm{Molina}, \binits{A.}}:
\batitle{Systematic mapping of digital gap and gender, age, ethnicity, or disability}.
\bjtitle{Sustainability}
\bvolume{14}(\bissue{3}),
\bfpage{1297}
(\byear{2022})
\end{barticle}
\endbibitem

\bibitem[\protect\citeauthoryear{Raja and Nagasubramani}{2018}]{raja2018impact}
\begin{barticle}
\bauthor{\bsnm{Raja}, \binits{R.}},
\bauthor{\bsnm{Nagasubramani}, \binits{P.}}:
\batitle{Impact of modern technology in education}.
\bjtitle{Journal of Applied and Advanced Research}
\bvolume{3}(\bissue{1}),
\bfpage{33}--\blpage{35}
(\byear{2018})
\end{barticle}
\endbibitem

\bibitem[\protect\citeauthoryear{Ben-Ari}{1998}]{ben1998constructivism}
\begin{barticle}
\bauthor{\bsnm{Ben-Ari}, \binits{M.}}:
\batitle{Constructivism in computer science education}.
\bjtitle{Acm sigcse bulletin}
\bvolume{30}(\bissue{1}),
\bfpage{257}--\blpage{261}
(\byear{1998})
\end{barticle}
\endbibitem

\bibitem[\protect\citeauthoryear{Sentance et~al.}{2017}]{sentance2017creating}
\begin{bchapter}
\bauthor{\bsnm{Sentance}, \binits{S.}},
\bauthor{\bsnm{Waite}, \binits{J.}},
\bauthor{\bsnm{Hodges}, \binits{S.}},
\bauthor{\bsnm{MacLeod}, \binits{E.}},
\bauthor{\bsnm{Yeomans}, \binits{L.}}:
\bctitle{" creating cool stuff" pupils' experience of the bbc micro: bit}.
In: \bbtitle{Proceedings of the 2017 ACM SIGCSE Technical Symposium on Computer Science Education},
pp. \bfpage{531}--\blpage{536}
(\byear{2017})
\end{bchapter}
\endbibitem

\bibitem[\protect\citeauthoryear{Fessard et~al.}{2019}]{fessard2019there}
\begin{bchapter}
\bauthor{\bsnm{Fessard}, \binits{G.}},
\bauthor{\bsnm{Wang}, \binits{P.}},
\bauthor{\bsnm{Renna}, \binits{I.}}:
\bctitle{Are there differences in learning gains when programming a tangible object or a simulation?}
In: \bbtitle{Proceedings of the 2019 ACM Conference on Innovation and Technology in Computer Science Education},
pp. \bfpage{78}--\blpage{84}
(\byear{2019})
\end{bchapter}
\endbibitem

\bibitem[\protect\citeauthoryear{Gressard and Loyd}{1985}]{gressard1985}
\begin{barticle}
\bauthor{\bsnm{Gressard}, \binits{C.}},
\bauthor{\bsnm{Loyd}, \binits{B.H.}}:
\batitle{Age and staff development experience with computers as factors affecting teacher attitudes toward computers}.
\bjtitle{School Science and Mathematics}
\bvolume{85}(\bissue{3}),
\bfpage{203}--\blpage{209}
(\byear{1985})
\doiurl{10.1111/j.1949-8594.1985.tb09613.x}
{\href{https://arxiv.org/abs/https://onlinelibrary.wiley.com/doi/pdf/10.1111/j.1949-8594.1985.tb09613.x}{{https://onlinelibrary.wiley.com/doi/pdf/10.1111/j.1949-8594.1985.tb09613.x}}}
\end{barticle}
\endbibitem

\bibitem[\protect\citeauthoryear{Butler and Sellbom}{2002}]{butler2002}
\begin{barticle}
\bauthor{\bsnm{Butler}, \binits{D.L.}},
\bauthor{\bsnm{Sellbom}, \binits{M.}}:
\batitle{Barriers to adopting technology for teaching and learning}.
\bjtitle{Educause Quarterly}
\bvolume{25}(\bissue{2}),
\bfpage{22}--\blpage{28}
(\byear{2002})
\end{barticle}
\endbibitem

\bibitem[\protect\citeauthoryear{Chizmar and Williams}{2001}]{chizmar2001}
\begin{barticle}
\bauthor{\bsnm{Chizmar}, \binits{J.F.}},
\bauthor{\bsnm{Williams}, \binits{D.B.}}:
\batitle{What do faculty want?}
\bjtitle{Educause Quarterly}
\bvolume{24}(\bissue{1}),
\bfpage{18}--\blpage{24}
(\byear{2001})
\end{barticle}
\endbibitem

\bibitem[\protect\citeauthoryear{Tinio}{2020}]{tinio2020}
\begin{botherref}
\oauthor{\bsnm{Tinio}, \binits{V.L.}}:
ICT in Education
(2020).
\url{https://api.semanticscholar.org/CorpusID:51763804}
\end{botherref}
\endbibitem

\bibitem[\protect\citeauthoryear{James and Engelhardt}{2012}]{james2012}
\begin{barticle}
\bauthor{\bsnm{James}, \binits{K.H.}},
\bauthor{\bsnm{Engelhardt}, \binits{L.}}:
\batitle{The effects of handwriting experience on functional brain development in pre-literate children}.
\bjtitle{Trends in Neuroscience and Education}
\bvolume{1}(\bissue{1}),
\bfpage{32}--\blpage{42}
(\byear{2012})
\doiurl{10.1016/j.tine.2012.08.001}
\end{barticle}
\endbibitem

\bibitem[\protect\citeauthoryear{Mayer et~al.}{2020}]{mayer2020}
\begin{barticle}
\bauthor{\bsnm{Mayer}, \binits{C.}},
\bauthor{\bsnm{Wallner}, \binits{S.}},
\bauthor{\bsnm{Budde-Spengler}, \binits{N.}},
\bauthor{\bsnm{Braunert}, \binits{S.}},
\bauthor{\bsnm{Arndt}, \binits{P.A.}},
\bauthor{\bsnm{Kiefer}, \binits{M.}}:
\batitle{Literacy training of kindergarten children with pencil, keyboard or tablet stylus: The influence of the writing tool on reading and writing performance at the letter and word level}.
\bjtitle{Frontiers in psychology}
\bvolume{10},
\bfpage{3054}
(\byear{2020})
\end{barticle}
\endbibitem

\bibitem[\protect\citeauthoryear{Muppalla et~al.}{2023}]{muppalla2023effects}
\begin{botherref}
\oauthor{\bsnm{Muppalla}, \binits{S.K.}},
\oauthor{\bsnm{Vuppalapati}, \binits{S.}},
\oauthor{\bsnm{Pulliahgaru}, \binits{A.R.}},
\oauthor{\bsnm{Sreenivasulu}, \binits{H.}}:
Effects of excessive screen time on child development: an updated review and strategies for management.
Cureus
\textbf{15}(6)
(2023)
\end{botherref}
\endbibitem

\bibitem[\protect\citeauthoryear{Tadpatrikar et~al.}{2024}]{tadpatrikar2024digital}
\begin{barticle}
\bauthor{\bsnm{Tadpatrikar}, \binits{A.}},
\bauthor{\bsnm{Sharma}, \binits{M.K.}},
\bauthor{\bsnm{Murthy}, \binits{P.}}:
\batitle{Digital dilemmas and existing recommendations for healthy screen time use for children and adolescents}.
\bjtitle{Asian Journal of Psychiatry}
\bvolume{99},
\bfpage{104154}
(\byear{2024})
\end{barticle}
\endbibitem

\bibitem[\protect\citeauthoryear{Al~Husaeni et~al.}{2024}]{al2024technology}
\begin{barticle}
\bauthor{\bsnm{Al~Husaeni}, \binits{D.F.}},
\bauthor{\bsnm{Al~Husaeni}, \binits{D.N.}},
\bauthor{\bsnm{Nandiyanto}, \binits{A.B.D.}},
\bauthor{\bsnm{Rokhman}, \binits{M.}},
\bauthor{\bsnm{Chalim}, \binits{S.}},
\bauthor{\bsnm{Chano}, \binits{J.}},
\bauthor{\bsnm{Al~Obaidi}, \binits{A.S.M.}},
\bauthor{\bsnm{Roestamy}, \binits{M.}}:
\batitle{How technology can change educational research? definition, factors for improving quality of education and computational bibliometric analysis}.
\bjtitle{ASEAN Journal of Science and Engineering}
\bvolume{4}(\bissue{2}),
\bfpage{127}--\blpage{166}
(\byear{2024})
\end{barticle}
\endbibitem

\bibitem[\protect\citeauthoryear{Ccelik and Baturay}{2024}]{ccelik2024technology}
\begin{botherref}
\oauthor{\bsnm{Ccelik}, \binits{F.}},
\oauthor{\bsnm{Baturay}, \binits{M.H.}}:
Technology and innovation in shaping the future of education.
Springer
(2024)
\end{botherref}
\endbibitem

\bibitem[\protect\citeauthoryear{Breuer and Bente}{2010}]{breuer2010so}
\begin{barticle}
\bauthor{\bsnm{Breuer}, \binits{J.}},
\bauthor{\bsnm{Bente}, \binits{G.}}:
\batitle{Why so serious? on the relation of serious games and learning}.
\bjtitle{Journal for computer game culture}
\bvolume{4},
\bfpage{7}--\blpage{24}
(\byear{2010})
\end{barticle}
\endbibitem

\bibitem[\protect\citeauthoryear{Shannon}{1951}]{shannon1950}
\begin{barticle}
\bauthor{\bsnm{Shannon}, \binits{C.E.}}:
\batitle{Prediction and entropy of printed english}.
\bjtitle{The Bell System Technical Journal}
\bvolume{30}(\bissue{1}),
\bfpage{50}--\blpage{64}
(\byear{1951})
\doiurl{10.1002/j.1538-7305.1951.tb01366.x}
\end{barticle}
\endbibitem

\bibitem[\protect\citeauthoryear{Zhai et~al.}{2008}]{zhai2008}
\begin{barticle}
\bauthor{\bsnm{Zhai}, \binits{C.}}, \betal:
\batitle{Statistical language models for information retrieval a critical review}.
\bjtitle{Foundations and Trends{\textregistered} in Information Retrieval}
\bvolume{2}(\bissue{3}),
\bfpage{137}--\blpage{213}
(\byear{2008})
\end{barticle}
\endbibitem

\bibitem[\protect\citeauthoryear{Chen and Goodman}{1999}]{CHEN1999359}
\begin{barticle}
\bauthor{\bsnm{Chen}, \binits{S.F.}},
\bauthor{\bsnm{Goodman}, \binits{J.}}:
\batitle{An empirical study of smoothing techniques for language modeling}.
\bjtitle{Computer Speech \& Language}
\bvolume{13}(\bissue{4}),
\bfpage{359}--\blpage{394}
(\byear{1999})
\end{barticle}
\endbibitem

\bibitem[\protect\citeauthoryear{Bengio et~al.}{2000}]{bengio2000neural}
\begin{botherref}
\oauthor{\bsnm{Bengio}, \binits{Y.}},
\oauthor{\bsnm{Ducharme}, \binits{R.}},
\oauthor{\bsnm{Vincent}, \binits{P.}}:
A neural probabilistic language model.
Advances in neural information processing systems
\textbf{13}
(2000)
\end{botherref}
\endbibitem

\bibitem[\protect\citeauthoryear{Devlin}{2018}]{devlin2018bert}
\begin{botherref}
\oauthor{\bsnm{Devlin}, \binits{J.}}:
Bert: Pre-training of deep bidirectional transformers for language understanding
(2018)
\end{botherref}
\endbibitem

\bibitem[\protect\citeauthoryear{Radford}{2018}]{radford2018improving}
\begin{botherref}
\oauthor{\bsnm{Radford}, \binits{A.}}:
Improving language understanding by generative pre-training
(2018)
\end{botherref}
\endbibitem

\bibitem[\protect\citeauthoryear{Zhou et~al.}{2024}]{zhou2024comprehensive}
\begin{botherref}
\oauthor{\bsnm{Zhou}, \binits{C.}},
\oauthor{\bsnm{Li}, \binits{Q.}},
\oauthor{\bsnm{Li}, \binits{C.}},
\oauthor{\bsnm{Yu}, \binits{J.}},
\oauthor{\bsnm{Liu}, \binits{Y.}},
\oauthor{\bsnm{Wang}, \binits{G.}},
\oauthor{\bsnm{Zhang}, \binits{K.}},
\oauthor{\bsnm{Ji}, \binits{C.}},
\oauthor{\bsnm{Yan}, \binits{Q.}},
\oauthor{\bsnm{He}, \binits{L.}}, et al.:
A comprehensive survey on pretrained foundation models: A history from bert to chatgpt.
Springer
(2024)
\end{botherref}
\endbibitem

\bibitem[\protect\citeauthoryear{Grattafiori et~al.}{2024}]{grattafiori2024llama}
\begin{botherref}
\oauthor{\bsnm{Grattafiori}, \binits{A.}},
\oauthor{\bsnm{Dubey}, \binits{A.}},
\oauthor{\bsnm{Jauhri}, \binits{A.}},
\oauthor{\bsnm{Pandey}, \binits{A.}},
\oauthor{\bsnm{Kadian}, \binits{A.}},
\oauthor{\bsnm{Al-Dahle}, \binits{A.}},
\oauthor{\bsnm{Letman}, \binits{A.}},
\oauthor{\bsnm{Mathur}, \binits{A.}},
\oauthor{\bsnm{Schelten}, \binits{A.}},
\oauthor{\bsnm{Vaughan}, \binits{A.}}, et al.:
The llama 3 herd of models.
arXiv e-prints,
2407
(2024)
\end{botherref}
\endbibitem

\bibitem[\protect\citeauthoryear{Jiang et~al.}{2024}]{jiang2024mixtralexperts}
\begin{botherref}
\oauthor{\bsnm{Jiang}, \binits{A.Q.}},
\oauthor{\bsnm{Sablayrolles}, \binits{A.}},
\oauthor{\bsnm{Roux}, \binits{A.}},
\oauthor{\bsnm{Mensch}, \binits{A.}},
\oauthor{\bsnm{Savary}, \binits{B.}},
\oauthor{\bsnm{Bamford}, \binits{C.}},
\oauthor{\bsnm{Chaplot}, \binits{D.S.}},
\oauthor{\bsnm{Casas}, \binits{D.}},
\oauthor{\bsnm{Hanna}, \binits{E.B.}},
\oauthor{\bsnm{Bressand}, \binits{F.}},
\oauthor{\bsnm{Lengyel}, \binits{G.}},
\oauthor{\bsnm{Bour}, \binits{G.}},
\oauthor{\bsnm{Lample}, \binits{G.}},
\oauthor{\bsnm{Lavaud}, \binits{L.R.}},
\oauthor{\bsnm{Saulnier}, \binits{L.}},
\oauthor{\bsnm{Lachaux}, \binits{M.-A.}},
\oauthor{\bsnm{Stock}, \binits{P.}},
\oauthor{\bsnm{Subramanian}, \binits{S.}},
\oauthor{\bsnm{Yang}, \binits{S.}},
\oauthor{\bsnm{Antoniak}, \binits{S.}},
\oauthor{\bsnm{Scao}, \binits{T.L.}},
\oauthor{\bsnm{Gervet}, \binits{T.}},
\oauthor{\bsnm{Lavril}, \binits{T.}},
\oauthor{\bsnm{Wang}, \binits{T.}},
\oauthor{\bsnm{Lacroix}, \binits{T.}},
\oauthor{\bsnm{Sayed}, \binits{W.E.}}:
Mixtral of Experts
(2024).
\url{https://arxiv.org/abs/2401.04088}
\end{botherref}
\endbibitem

\bibitem[\protect\citeauthoryear{Team et~al.}{2024}]{gemmateam2024gemmaopenmodelsbased}
\begin{botherref}
\oauthor{\bsnm{Team}, \binits{G.}},
\oauthor{\bsnm{Mesnard}, \binits{T.}},
\oauthor{\bsnm{Hardin}, \binits{C.}},
\oauthor{\bsnm{Dadashi}, \binits{R.}},
\oauthor{\bsnm{Bhupatiraju}, \binits{S.}},
\oauthor{\bsnm{Pathak}, \binits{S.}},
\oauthor{\bsnm{Sifre}, \binits{L.}},
\oauthor{\bsnm{Rivière}, \binits{M.}},
\oauthor{\bsnm{Kale}, \binits{M.S.}},
\oauthor{\bsnm{Love}, \binits{J.}},
\oauthor{\bsnm{Tafti}, \binits{P.}},
\oauthor{\bsnm{Hussenot}, \binits{L.}},
\oauthor{\bsnm{Sessa}, \binits{P.G.}},
\oauthor{\bsnm{Chowdhery}, \binits{A.}},
\oauthor{\bsnm{Roberts}, \binits{A.}},
\oauthor{\bsnm{Barua}, \binits{A.}},
\oauthor{\bsnm{Botev}, \binits{A.}},
\oauthor{\bsnm{Castro-Ros}, \binits{A.}},
\oauthor{\bsnm{Slone}, \binits{A.}},
\oauthor{\bsnm{Héliou}, \binits{A.}},
\oauthor{\bsnm{Tacchetti}, \binits{A.}},
\oauthor{\bsnm{Bulanova}, \binits{A.}},
\oauthor{\bsnm{Paterson}, \binits{A.}},
\oauthor{\bsnm{Tsai}, \binits{B.}},
\oauthor{\bsnm{Shahriari}, \binits{B.}},
\oauthor{\bsnm{Lan}, \binits{C.L.}},
\oauthor{\bsnm{Choquette-Choo}, \binits{C.A.}},
\oauthor{\bsnm{Crepy}, \binits{C.}},
\oauthor{\bsnm{Cer}, \binits{D.}},
\oauthor{\bsnm{Ippolito}, \binits{D.}},
\oauthor{\bsnm{Reid}, \binits{D.}},
\oauthor{\bsnm{Buchatskaya}, \binits{E.}},
\oauthor{\bsnm{Ni}, \binits{E.}},
\oauthor{\bsnm{Noland}, \binits{E.}},
\oauthor{\bsnm{Yan}, \binits{G.}},
\oauthor{\bsnm{Tucker}, \binits{G.}},
\oauthor{\bsnm{Muraru}, \binits{G.-C.}},
\oauthor{\bsnm{Rozhdestvenskiy}, \binits{G.}},
\oauthor{\bsnm{Michalewski}, \binits{H.}},
\oauthor{\bsnm{Tenney}, \binits{I.}},
\oauthor{\bsnm{Grishchenko}, \binits{I.}},
\oauthor{\bsnm{Austin}, \binits{J.}},
\oauthor{\bsnm{Keeling}, \binits{J.}},
\oauthor{\bsnm{Labanowski}, \binits{J.}},
\oauthor{\bsnm{Lespiau}, \binits{J.-B.}},
\oauthor{\bsnm{Stanway}, \binits{J.}},
\oauthor{\bsnm{Brennan}, \binits{J.}},
\oauthor{\bsnm{Chen}, \binits{J.}},
\oauthor{\bsnm{Ferret}, \binits{J.}},
\oauthor{\bsnm{Chiu}, \binits{J.}},
\oauthor{\bsnm{Mao-Jones}, \binits{J.}},
\oauthor{\bsnm{Lee}, \binits{K.}},
\oauthor{\bsnm{Yu}, \binits{K.}},
\oauthor{\bsnm{Millican}, \binits{K.}},
\oauthor{\bsnm{Sjoesund}, \binits{L.L.}},
\oauthor{\bsnm{Lee}, \binits{L.}},
\oauthor{\bsnm{Dixon}, \binits{L.}},
\oauthor{\bsnm{Reid}, \binits{M.}},
\oauthor{\bsnm{Mikuła}, \binits{M.}},
\oauthor{\bsnm{Wirth}, \binits{M.}},
\oauthor{\bsnm{Sharman}, \binits{M.}},
\oauthor{\bsnm{Chinaev}, \binits{N.}},
\oauthor{\bsnm{Thain}, \binits{N.}},
\oauthor{\bsnm{Bachem}, \binits{O.}},
\oauthor{\bsnm{Chang}, \binits{O.}},
\oauthor{\bsnm{Wahltinez}, \binits{O.}},
\oauthor{\bsnm{Bailey}, \binits{P.}},
\oauthor{\bsnm{Michel}, \binits{P.}},
\oauthor{\bsnm{Yotov}, \binits{P.}},
\oauthor{\bsnm{Chaabouni}, \binits{R.}},
\oauthor{\bsnm{Comanescu}, \binits{R.}},
\oauthor{\bsnm{Jana}, \binits{R.}},
\oauthor{\bsnm{Anil}, \binits{R.}},
\oauthor{\bsnm{McIlroy}, \binits{R.}},
\oauthor{\bsnm{Liu}, \binits{R.}},
\oauthor{\bsnm{Mullins}, \binits{R.}},
\oauthor{\bsnm{Smith}, \binits{S.L.}},
\oauthor{\bsnm{Borgeaud}, \binits{S.}},
\oauthor{\bsnm{Girgin}, \binits{S.}},
\oauthor{\bsnm{Douglas}, \binits{S.}},
\oauthor{\bsnm{Pandya}, \binits{S.}},
\oauthor{\bsnm{Shakeri}, \binits{S.}},
\oauthor{\bsnm{De}, \binits{S.}},
\oauthor{\bsnm{Klimenko}, \binits{T.}},
\oauthor{\bsnm{Hennigan}, \binits{T.}},
\oauthor{\bsnm{Feinberg}, \binits{V.}},
\oauthor{\bsnm{Stokowiec}, \binits{W.}},
\oauthor{\bsnm{Chen}, \binits{Y.-h.}},
\oauthor{\bsnm{Ahmed}, \binits{Z.}},
\oauthor{\bsnm{Gong}, \binits{Z.}},
\oauthor{\bsnm{Warkentin}, \binits{T.}},
\oauthor{\bsnm{Peran}, \binits{L.}},
\oauthor{\bsnm{Giang}, \binits{M.}},
\oauthor{\bsnm{Farabet}, \binits{C.}},
\oauthor{\bsnm{Vinyals}, \binits{O.}},
\oauthor{\bsnm{Dean}, \binits{J.}},
\oauthor{\bsnm{Kavukcuoglu}, \binits{K.}},
\oauthor{\bsnm{Hassabis}, \binits{D.}},
\oauthor{\bsnm{Ghahramani}, \binits{Z.}},
\oauthor{\bsnm{Eck}, \binits{D.}},
\oauthor{\bsnm{Barral}, \binits{J.}},
\oauthor{\bsnm{Pereira}, \binits{F.}},
\oauthor{\bsnm{Collins}, \binits{E.}},
\oauthor{\bsnm{Joulin}, \binits{A.}},
\oauthor{\bsnm{Fiedel}, \binits{N.}},
\oauthor{\bsnm{Senter}, \binits{E.}},
\oauthor{\bsnm{Andreev}, \binits{A.}},
\oauthor{\bsnm{Kenealy}, \binits{K.}}:
Gemma: Open Models Based on Gemini Research and Technology
(2024).
\url{https://arxiv.org/abs/2403.08295}
\end{botherref}
\endbibitem

\bibitem[\protect\citeauthoryear{Yang et~al.}{2024}]{yang2024qwen2technicalreport}
\begin{botherref}
\oauthor{\bsnm{Yang}, \binits{A.}},
\oauthor{\bsnm{Yang}, \binits{B.}},
\oauthor{\bsnm{Hui}, \binits{B.}},
\oauthor{\bsnm{Zheng}, \binits{B.}},
\oauthor{\bsnm{Yu}, \binits{B.}},
\oauthor{\bsnm{Zhou}, \binits{C.}},
\oauthor{\bsnm{Li}, \binits{C.}},
\oauthor{\bsnm{Li}, \binits{C.}},
\oauthor{\bsnm{Liu}, \binits{D.}},
\oauthor{\bsnm{Huang}, \binits{F.}},
\oauthor{\bsnm{Dong}, \binits{G.}},
\oauthor{\bsnm{Wei}, \binits{H.}},
\oauthor{\bsnm{Lin}, \binits{H.}},
\oauthor{\bsnm{Tang}, \binits{J.}},
\oauthor{\bsnm{Wang}, \binits{J.}},
\oauthor{\bsnm{Yang}, \binits{J.}},
\oauthor{\bsnm{Tu}, \binits{J.}},
\oauthor{\bsnm{Zhang}, \binits{J.}},
\oauthor{\bsnm{Ma}, \binits{J.}},
\oauthor{\bsnm{Yang}, \binits{J.}},
\oauthor{\bsnm{Xu}, \binits{J.}},
\oauthor{\bsnm{Zhou}, \binits{J.}},
\oauthor{\bsnm{Bai}, \binits{J.}},
\oauthor{\bsnm{He}, \binits{J.}},
\oauthor{\bsnm{Lin}, \binits{J.}},
\oauthor{\bsnm{Dang}, \binits{K.}},
\oauthor{\bsnm{Lu}, \binits{K.}},
\oauthor{\bsnm{Chen}, \binits{K.}},
\oauthor{\bsnm{Yang}, \binits{K.}},
\oauthor{\bsnm{Li}, \binits{M.}},
\oauthor{\bsnm{Xue}, \binits{M.}},
\oauthor{\bsnm{Ni}, \binits{N.}},
\oauthor{\bsnm{Zhang}, \binits{P.}},
\oauthor{\bsnm{Wang}, \binits{P.}},
\oauthor{\bsnm{Peng}, \binits{R.}},
\oauthor{\bsnm{Men}, \binits{R.}},
\oauthor{\bsnm{Gao}, \binits{R.}},
\oauthor{\bsnm{Lin}, \binits{R.}},
\oauthor{\bsnm{Wang}, \binits{S.}},
\oauthor{\bsnm{Bai}, \binits{S.}},
\oauthor{\bsnm{Tan}, \binits{S.}},
\oauthor{\bsnm{Zhu}, \binits{T.}},
\oauthor{\bsnm{Li}, \binits{T.}},
\oauthor{\bsnm{Liu}, \binits{T.}},
\oauthor{\bsnm{Ge}, \binits{W.}},
\oauthor{\bsnm{Deng}, \binits{X.}},
\oauthor{\bsnm{Zhou}, \binits{X.}},
\oauthor{\bsnm{Ren}, \binits{X.}},
\oauthor{\bsnm{Zhang}, \binits{X.}},
\oauthor{\bsnm{Wei}, \binits{X.}},
\oauthor{\bsnm{Ren}, \binits{X.}},
\oauthor{\bsnm{Liu}, \binits{X.}},
\oauthor{\bsnm{Fan}, \binits{Y.}},
\oauthor{\bsnm{Yao}, \binits{Y.}},
\oauthor{\bsnm{Zhang}, \binits{Y.}},
\oauthor{\bsnm{Wan}, \binits{Y.}},
\oauthor{\bsnm{Chu}, \binits{Y.}},
\oauthor{\bsnm{Liu}, \binits{Y.}},
\oauthor{\bsnm{Cui}, \binits{Z.}},
\oauthor{\bsnm{Zhang}, \binits{Z.}},
\oauthor{\bsnm{Guo}, \binits{Z.}},
\oauthor{\bsnm{Fan}, \binits{Z.}}:
Qwen2 Technical Report
(2024).
\url{https://arxiv.org/abs/2407.10671}
\end{botherref}
\endbibitem

\bibitem[\protect\citeauthoryear{Chowdhery et~al.}{2022}]{chowdhery2022palmscalinglanguagemodeling}
\begin{botherref}
\oauthor{\bsnm{Chowdhery}, \binits{A.}},
\oauthor{\bsnm{Narang}, \binits{S.}},
\oauthor{\bsnm{Devlin}, \binits{J.}},
\oauthor{\bsnm{Bosma}, \binits{M.}},
\oauthor{\bsnm{Mishra}, \binits{G.}},
\oauthor{\bsnm{Roberts}, \binits{A.}},
\oauthor{\bsnm{Barham}, \binits{P.}},
\oauthor{\bsnm{Chung}, \binits{H.W.}},
\oauthor{\bsnm{Sutton}, \binits{C.}},
\oauthor{\bsnm{Gehrmann}, \binits{S.}},
\oauthor{\bsnm{Schuh}, \binits{P.}},
\oauthor{\bsnm{Shi}, \binits{K.}},
\oauthor{\bsnm{Tsvyashchenko}, \binits{S.}},
\oauthor{\bsnm{Maynez}, \binits{J.}},
\oauthor{\bsnm{Rao}, \binits{A.}},
\oauthor{\bsnm{Barnes}, \binits{P.}},
\oauthor{\bsnm{Tay}, \binits{Y.}},
\oauthor{\bsnm{Shazeer}, \binits{N.}},
\oauthor{\bsnm{Prabhakaran}, \binits{V.}},
\oauthor{\bsnm{Reif}, \binits{E.}},
\oauthor{\bsnm{Du}, \binits{N.}},
\oauthor{\bsnm{Hutchinson}, \binits{B.}},
\oauthor{\bsnm{Pope}, \binits{R.}},
\oauthor{\bsnm{Bradbury}, \binits{J.}},
\oauthor{\bsnm{Austin}, \binits{J.}},
\oauthor{\bsnm{Isard}, \binits{M.}},
\oauthor{\bsnm{Gur-Ari}, \binits{G.}},
\oauthor{\bsnm{Yin}, \binits{P.}},
\oauthor{\bsnm{Duke}, \binits{T.}},
\oauthor{\bsnm{Levskaya}, \binits{A.}},
\oauthor{\bsnm{Ghemawat}, \binits{S.}},
\oauthor{\bsnm{Dev}, \binits{S.}},
\oauthor{\bsnm{Michalewski}, \binits{H.}},
\oauthor{\bsnm{Garcia}, \binits{X.}},
\oauthor{\bsnm{Misra}, \binits{V.}},
\oauthor{\bsnm{Robinson}, \binits{K.}},
\oauthor{\bsnm{Fedus}, \binits{L.}},
\oauthor{\bsnm{Zhou}, \binits{D.}},
\oauthor{\bsnm{Ippolito}, \binits{D.}},
\oauthor{\bsnm{Luan}, \binits{D.}},
\oauthor{\bsnm{Lim}, \binits{H.}},
\oauthor{\bsnm{Zoph}, \binits{B.}},
\oauthor{\bsnm{Spiridonov}, \binits{A.}},
\oauthor{\bsnm{Sepassi}, \binits{R.}},
\oauthor{\bsnm{Dohan}, \binits{D.}},
\oauthor{\bsnm{Agrawal}, \binits{S.}},
\oauthor{\bsnm{Omernick}, \binits{M.}},
\oauthor{\bsnm{Dai}, \binits{A.M.}},
\oauthor{\bsnm{Pillai}, \binits{T.S.}},
\oauthor{\bsnm{Pellat}, \binits{M.}},
\oauthor{\bsnm{Lewkowycz}, \binits{A.}},
\oauthor{\bsnm{Moreira}, \binits{E.}},
\oauthor{\bsnm{Child}, \binits{R.}},
\oauthor{\bsnm{Polozov}, \binits{O.}},
\oauthor{\bsnm{Lee}, \binits{K.}},
\oauthor{\bsnm{Zhou}, \binits{Z.}},
\oauthor{\bsnm{Wang}, \binits{X.}},
\oauthor{\bsnm{Saeta}, \binits{B.}},
\oauthor{\bsnm{Diaz}, \binits{M.}},
\oauthor{\bsnm{Firat}, \binits{O.}},
\oauthor{\bsnm{Catasta}, \binits{M.}},
\oauthor{\bsnm{Wei}, \binits{J.}},
\oauthor{\bsnm{Meier-Hellstern}, \binits{K.}},
\oauthor{\bsnm{Eck}, \binits{D.}},
\oauthor{\bsnm{Dean}, \binits{J.}},
\oauthor{\bsnm{Petrov}, \binits{S.}},
\oauthor{\bsnm{Fiedel}, \binits{N.}}:
PaLM: Scaling Language Modeling with Pathways
(2022).
\url{https://arxiv.org/abs/2204.02311}
\end{botherref}
\endbibitem

\bibitem[\protect\citeauthoryear{OpenAI et~al.}{2024}]{openai2024gpt4technicalreport}
\begin{botherref}
\oauthor{\bsnm{OpenAI}},
\oauthor{\bsnm{Achiam}, \binits{J.}},
\oauthor{\bsnm{Adler}, \binits{S.}},
\oauthor{\bsnm{Agarwal}, \binits{S.}},
\oauthor{\bsnm{Ahmad}, \binits{L.}},
\oauthor{\bsnm{Akkaya}, \binits{I.}},
\oauthor{\bsnm{Aleman}, \binits{F.L.}},
\oauthor{\bsnm{Almeida}, \binits{D.}},
\oauthor{\bsnm{Altenschmidt}, \binits{J.}},
\oauthor{\bsnm{Altman}, \binits{S.}},
\oauthor{\bsnm{Anadkat}, \binits{S.}},
\oauthor{\bsnm{Avila}, \binits{R.}},
\oauthor{\bsnm{Babuschkin}, \binits{I.}},
\oauthor{\bsnm{Balaji}, \binits{S.}},
\oauthor{\bsnm{Balcom}, \binits{V.}},
\oauthor{\bsnm{Baltescu}, \binits{P.}},
\oauthor{\bsnm{Bao}, \binits{H.}},
\oauthor{\bsnm{Bavarian}, \binits{M.}},
\oauthor{\bsnm{Belgum}, \binits{J.}},
\oauthor{\bsnm{Bello}, \binits{I.}},
\oauthor{\bsnm{Berdine}, \binits{J.}},
\oauthor{\bsnm{Bernadett-Shapiro}, \binits{G.}},
\oauthor{\bsnm{Berner}, \binits{C.}},
\oauthor{\bsnm{Bogdonoff}, \binits{L.}},
\oauthor{\bsnm{Boiko}, \binits{O.}},
\oauthor{\bsnm{Boyd}, \binits{M.}},
\oauthor{\bsnm{Brakman}, \binits{A.-L.}},
\oauthor{\bsnm{Brockman}, \binits{G.}},
\oauthor{\bsnm{Brooks}, \binits{T.}},
\oauthor{\bsnm{Brundage}, \binits{M.}},
\oauthor{\bsnm{Button}, \binits{K.}},
\oauthor{\bsnm{Cai}, \binits{T.}},
\oauthor{\bsnm{Campbell}, \binits{R.}},
\oauthor{\bsnm{Cann}, \binits{A.}},
\oauthor{\bsnm{Carey}, \binits{B.}},
\oauthor{\bsnm{Carlson}, \binits{C.}},
\oauthor{\bsnm{Carmichael}, \binits{R.}},
\oauthor{\bsnm{Chan}, \binits{B.}},
\oauthor{\bsnm{Chang}, \binits{C.}},
\oauthor{\bsnm{Chantzis}, \binits{F.}},
\oauthor{\bsnm{Chen}, \binits{D.}},
\oauthor{\bsnm{Chen}, \binits{S.}},
\oauthor{\bsnm{Chen}, \binits{R.}},
\oauthor{\bsnm{Chen}, \binits{J.}},
\oauthor{\bsnm{Chen}, \binits{M.}},
\oauthor{\bsnm{Chess}, \binits{B.}},
\oauthor{\bsnm{Cho}, \binits{C.}},
\oauthor{\bsnm{Chu}, \binits{C.}},
\oauthor{\bsnm{Chung}, \binits{H.W.}},
\oauthor{\bsnm{Cummings}, \binits{D.}},
\oauthor{\bsnm{Currier}, \binits{J.}},
\oauthor{\bsnm{Dai}, \binits{Y.}},
\oauthor{\bsnm{Decareaux}, \binits{C.}},
\oauthor{\bsnm{Degry}, \binits{T.}},
\oauthor{\bsnm{Deutsch}, \binits{N.}},
\oauthor{\bsnm{Deville}, \binits{D.}},
\oauthor{\bsnm{Dhar}, \binits{A.}},
\oauthor{\bsnm{Dohan}, \binits{D.}},
\oauthor{\bsnm{Dowling}, \binits{S.}},
\oauthor{\bsnm{Dunning}, \binits{S.}},
\oauthor{\bsnm{Ecoffet}, \binits{A.}},
\oauthor{\bsnm{Eleti}, \binits{A.}},
\oauthor{\bsnm{Eloundou}, \binits{T.}},
\oauthor{\bsnm{Farhi}, \binits{D.}},
\oauthor{\bsnm{Fedus}, \binits{L.}},
\oauthor{\bsnm{Felix}, \binits{N.}},
\oauthor{\bsnm{Fishman}, \binits{S.P.}},
\oauthor{\bsnm{Forte}, \binits{J.}},
\oauthor{\bsnm{Fulford}, \binits{I.}},
\oauthor{\bsnm{Gao}, \binits{L.}},
\oauthor{\bsnm{Georges}, \binits{E.}},
\oauthor{\bsnm{Gibson}, \binits{C.}},
\oauthor{\bsnm{Goel}, \binits{V.}},
\oauthor{\bsnm{Gogineni}, \binits{T.}},
\oauthor{\bsnm{Goh}, \binits{G.}},
\oauthor{\bsnm{Gontijo-Lopes}, \binits{R.}},
\oauthor{\bsnm{Gordon}, \binits{J.}},
\oauthor{\bsnm{Grafstein}, \binits{M.}},
\oauthor{\bsnm{Gray}, \binits{S.}},
\oauthor{\bsnm{Greene}, \binits{R.}},
\oauthor{\bsnm{Gross}, \binits{J.}},
\oauthor{\bsnm{Gu}, \binits{S.S.}},
\oauthor{\bsnm{Guo}, \binits{Y.}},
\oauthor{\bsnm{Hallacy}, \binits{C.}},
\oauthor{\bsnm{Han}, \binits{J.}},
\oauthor{\bsnm{Harris}, \binits{J.}},
\oauthor{\bsnm{He}, \binits{Y.}},
\oauthor{\bsnm{Heaton}, \binits{M.}},
\oauthor{\bsnm{Heidecke}, \binits{J.}},
\oauthor{\bsnm{Hesse}, \binits{C.}},
\oauthor{\bsnm{Hickey}, \binits{A.}},
\oauthor{\bsnm{Hickey}, \binits{W.}},
\oauthor{\bsnm{Hoeschele}, \binits{P.}},
\oauthor{\bsnm{Houghton}, \binits{B.}},
\oauthor{\bsnm{Hsu}, \binits{K.}},
\oauthor{\bsnm{Hu}, \binits{S.}},
\oauthor{\bsnm{Hu}, \binits{X.}},
\oauthor{\bsnm{Huizinga}, \binits{J.}},
\oauthor{\bsnm{Jain}, \binits{S.}},
\oauthor{\bsnm{Jain}, \binits{S.}},
\oauthor{\bsnm{Jang}, \binits{J.}},
\oauthor{\bsnm{Jiang}, \binits{A.}},
\oauthor{\bsnm{Jiang}, \binits{R.}},
\oauthor{\bsnm{Jin}, \binits{H.}},
\oauthor{\bsnm{Jin}, \binits{D.}},
\oauthor{\bsnm{Jomoto}, \binits{S.}},
\oauthor{\bsnm{Jonn}, \binits{B.}},
\oauthor{\bsnm{Jun}, \binits{H.}},
\oauthor{\bsnm{Kaftan}, \binits{T.}},
\oauthor{\bsnm{Kaiser}},
\oauthor{\bsnm{Kamali}, \binits{A.}},
\oauthor{\bsnm{Kanitscheider}, \binits{I.}},
\oauthor{\bsnm{Keskar}, \binits{N.S.}},
\oauthor{\bsnm{Khan}, \binits{T.}},
\oauthor{\bsnm{Kilpatrick}, \binits{L.}},
\oauthor{\bsnm{Kim}, \binits{J.W.}},
\oauthor{\bsnm{Kim}, \binits{C.}},
\oauthor{\bsnm{Kim}, \binits{Y.}},
\oauthor{\bsnm{Kirchner}, \binits{J.H.}},
\oauthor{\bsnm{Kiros}, \binits{J.}},
\oauthor{\bsnm{Knight}, \binits{M.}},
\oauthor{\bsnm{Kokotajlo}, \binits{D.}},
\oauthor{\bsnm{Kondraciuk}},
\oauthor{\bsnm{Kondrich}, \binits{A.}},
\oauthor{\bsnm{Konstantinidis}, \binits{A.}},
\oauthor{\bsnm{Kosic}, \binits{K.}},
\oauthor{\bsnm{Krueger}, \binits{G.}},
\oauthor{\bsnm{Kuo}, \binits{V.}},
\oauthor{\bsnm{Lampe}, \binits{M.}},
\oauthor{\bsnm{Lan}, \binits{I.}},
\oauthor{\bsnm{Lee}, \binits{T.}},
\oauthor{\bsnm{Leike}, \binits{J.}},
\oauthor{\bsnm{Leung}, \binits{J.}},
\oauthor{\bsnm{Levy}, \binits{D.}},
\oauthor{\bsnm{Li}, \binits{C.M.}},
\oauthor{\bsnm{Lim}, \binits{R.}},
\oauthor{\bsnm{Lin}, \binits{M.}},
\oauthor{\bsnm{Lin}, \binits{S.}},
\oauthor{\bsnm{Litwin}, \binits{M.}},
\oauthor{\bsnm{Lopez}, \binits{T.}},
\oauthor{\bsnm{Lowe}, \binits{R.}},
\oauthor{\bsnm{Lue}, \binits{P.}},
\oauthor{\bsnm{Makanju}, \binits{A.}},
\oauthor{\bsnm{Malfacini}, \binits{K.}},
\oauthor{\bsnm{Manning}, \binits{S.}},
\oauthor{\bsnm{Markov}, \binits{T.}},
\oauthor{\bsnm{Markovski}, \binits{Y.}},
\oauthor{\bsnm{Martin}, \binits{B.}},
\oauthor{\bsnm{Mayer}, \binits{K.}},
\oauthor{\bsnm{Mayne}, \binits{A.}},
\oauthor{\bsnm{McGrew}, \binits{B.}},
\oauthor{\bsnm{McKinney}, \binits{S.M.}},
\oauthor{\bsnm{McLeavey}, \binits{C.}},
\oauthor{\bsnm{McMillan}, \binits{P.}},
\oauthor{\bsnm{McNeil}, \binits{J.}},
\oauthor{\bsnm{Medina}, \binits{D.}},
\oauthor{\bsnm{Mehta}, \binits{A.}},
\oauthor{\bsnm{Menick}, \binits{J.}},
\oauthor{\bsnm{Metz}, \binits{L.}},
\oauthor{\bsnm{Mishchenko}, \binits{A.}},
\oauthor{\bsnm{Mishkin}, \binits{P.}},
\oauthor{\bsnm{Monaco}, \binits{V.}},
\oauthor{\bsnm{Morikawa}, \binits{E.}},
\oauthor{\bsnm{Mossing}, \binits{D.}},
\oauthor{\bsnm{Mu}, \binits{T.}},
\oauthor{\bsnm{Murati}, \binits{M.}},
\oauthor{\bsnm{Murk}, \binits{O.}},
\oauthor{\bsnm{Mély}, \binits{D.}},
\oauthor{\bsnm{Nair}, \binits{A.}},
\oauthor{\bsnm{Nakano}, \binits{R.}},
\oauthor{\bsnm{Nayak}, \binits{R.}},
\oauthor{\bsnm{Neelakantan}, \binits{A.}},
\oauthor{\bsnm{Ngo}, \binits{R.}},
\oauthor{\bsnm{Noh}, \binits{H.}},
\oauthor{\bsnm{Ouyang}, \binits{L.}},
\oauthor{\bsnm{O'Keefe}, \binits{C.}},
\oauthor{\bsnm{Pachocki}, \binits{J.}},
\oauthor{\bsnm{Paino}, \binits{A.}},
\oauthor{\bsnm{Palermo}, \binits{J.}},
\oauthor{\bsnm{Pantuliano}, \binits{A.}},
\oauthor{\bsnm{Parascandolo}, \binits{G.}},
\oauthor{\bsnm{Parish}, \binits{J.}},
\oauthor{\bsnm{Parparita}, \binits{E.}},
\oauthor{\bsnm{Passos}, \binits{A.}},
\oauthor{\bsnm{Pavlov}, \binits{M.}},
\oauthor{\bsnm{Peng}, \binits{A.}},
\oauthor{\bsnm{Perelman}, \binits{A.}},
\oauthor{\bsnm{Avila Belbute~Peres}, \binits{F.}},
\oauthor{\bsnm{Petrov}, \binits{M.}},
\oauthor{\bsnm{Oliveira~Pinto}, \binits{H.P.}},
\oauthor{\bsnm{Michael}},
\oauthor{\bsnm{Pokorny}},
\oauthor{\bsnm{Pokrass}, \binits{M.}},
\oauthor{\bsnm{Pong}, \binits{V.H.}},
\oauthor{\bsnm{Powell}, \binits{T.}},
\oauthor{\bsnm{Power}, \binits{A.}},
\oauthor{\bsnm{Power}, \binits{B.}},
\oauthor{\bsnm{Proehl}, \binits{E.}},
\oauthor{\bsnm{Puri}, \binits{R.}},
\oauthor{\bsnm{Radford}, \binits{A.}},
\oauthor{\bsnm{Rae}, \binits{J.}},
\oauthor{\bsnm{Ramesh}, \binits{A.}},
\oauthor{\bsnm{Raymond}, \binits{C.}},
\oauthor{\bsnm{Real}, \binits{F.}},
\oauthor{\bsnm{Rimbach}, \binits{K.}},
\oauthor{\bsnm{Ross}, \binits{C.}},
\oauthor{\bsnm{Rotsted}, \binits{B.}},
\oauthor{\bsnm{Roussez}, \binits{H.}},
\oauthor{\bsnm{Ryder}, \binits{N.}},
\oauthor{\bsnm{Saltarelli}, \binits{M.}},
\oauthor{\bsnm{Sanders}, \binits{T.}},
\oauthor{\bsnm{Santurkar}, \binits{S.}},
\oauthor{\bsnm{Sastry}, \binits{G.}},
\oauthor{\bsnm{Schmidt}, \binits{H.}},
\oauthor{\bsnm{Schnurr}, \binits{D.}},
\oauthor{\bsnm{Schulman}, \binits{J.}},
\oauthor{\bsnm{Selsam}, \binits{D.}},
\oauthor{\bsnm{Sheppard}, \binits{K.}},
\oauthor{\bsnm{Sherbakov}, \binits{T.}},
\oauthor{\bsnm{Shieh}, \binits{J.}},
\oauthor{\bsnm{Shoker}, \binits{S.}},
\oauthor{\bsnm{Shyam}, \binits{P.}},
\oauthor{\bsnm{Sidor}, \binits{S.}},
\oauthor{\bsnm{Sigler}, \binits{E.}},
\oauthor{\bsnm{Simens}, \binits{M.}},
\oauthor{\bsnm{Sitkin}, \binits{J.}},
\oauthor{\bsnm{Slama}, \binits{K.}},
\oauthor{\bsnm{Sohl}, \binits{I.}},
\oauthor{\bsnm{Sokolowsky}, \binits{B.}},
\oauthor{\bsnm{Song}, \binits{Y.}},
\oauthor{\bsnm{Staudacher}, \binits{N.}},
\oauthor{\bsnm{Such}, \binits{F.P.}},
\oauthor{\bsnm{Summers}, \binits{N.}},
\oauthor{\bsnm{Sutskever}, \binits{I.}},
\oauthor{\bsnm{Tang}, \binits{J.}},
\oauthor{\bsnm{Tezak}, \binits{N.}},
\oauthor{\bsnm{Thompson}, \binits{M.B.}},
\oauthor{\bsnm{Tillet}, \binits{P.}},
\oauthor{\bsnm{Tootoonchian}, \binits{A.}},
\oauthor{\bsnm{Tseng}, \binits{E.}},
\oauthor{\bsnm{Tuggle}, \binits{P.}},
\oauthor{\bsnm{Turley}, \binits{N.}},
\oauthor{\bsnm{Tworek}, \binits{J.}},
\oauthor{\bsnm{Uribe}, \binits{J.F.C.}},
\oauthor{\bsnm{Vallone}, \binits{A.}},
\oauthor{\bsnm{Vijayvergiya}, \binits{A.}},
\oauthor{\bsnm{Voss}, \binits{C.}},
\oauthor{\bsnm{Wainwright}, \binits{C.}},
\oauthor{\bsnm{Wang}, \binits{J.J.}},
\oauthor{\bsnm{Wang}, \binits{A.}},
\oauthor{\bsnm{Wang}, \binits{B.}},
\oauthor{\bsnm{Ward}, \binits{J.}},
\oauthor{\bsnm{Wei}, \binits{J.}},
\oauthor{\bsnm{Weinmann}, \binits{C.}},
\oauthor{\bsnm{Welihinda}, \binits{A.}},
\oauthor{\bsnm{Welinder}, \binits{P.}},
\oauthor{\bsnm{Weng}, \binits{J.}},
\oauthor{\bsnm{Weng}, \binits{L.}},
\oauthor{\bsnm{Wiethoff}, \binits{M.}},
\oauthor{\bsnm{Willner}, \binits{D.}},
\oauthor{\bsnm{Winter}, \binits{C.}},
\oauthor{\bsnm{Wolrich}, \binits{S.}},
\oauthor{\bsnm{Wong}, \binits{H.}},
\oauthor{\bsnm{Workman}, \binits{L.}},
\oauthor{\bsnm{Wu}, \binits{S.}},
\oauthor{\bsnm{Wu}, \binits{J.}},
\oauthor{\bsnm{Wu}, \binits{M.}},
\oauthor{\bsnm{Xiao}, \binits{K.}},
\oauthor{\bsnm{Xu}, \binits{T.}},
\oauthor{\bsnm{Yoo}, \binits{S.}},
\oauthor{\bsnm{Yu}, \binits{K.}},
\oauthor{\bsnm{Yuan}, \binits{Q.}},
\oauthor{\bsnm{Zaremba}, \binits{W.}},
\oauthor{\bsnm{Zellers}, \binits{R.}},
\oauthor{\bsnm{Zhang}, \binits{C.}},
\oauthor{\bsnm{Zhang}, \binits{M.}},
\oauthor{\bsnm{Zhao}, \binits{S.}},
\oauthor{\bsnm{Zheng}, \binits{T.}},
\oauthor{\bsnm{Zhuang}, \binits{J.}},
\oauthor{\bsnm{Zhuk}, \binits{W.}},
\oauthor{\bsnm{Zoph}, \binits{B.}}:
GPT-4 Technical Report
(2024).
\url{https://arxiv.org/abs/2303.08774}
\end{botherref}
\endbibitem

\bibitem[\protect\citeauthoryear{Zhao et~al.}{2023}]{zhao2024surveylargelanguagemodels}
\begin{botherref}
\oauthor{\bsnm{Zhao}, \binits{W.X.}},
\oauthor{\bsnm{Zhou}, \binits{K.}},
\oauthor{\bsnm{Li}, \binits{J.}},
\oauthor{\bsnm{Tang}, \binits{T.}},
\oauthor{\bsnm{Wang}, \binits{X.}},
\oauthor{\bsnm{Hou}, \binits{Y.}},
\oauthor{\bsnm{Min}, \binits{Y.}},
\oauthor{\bsnm{Zhang}, \binits{B.}},
\oauthor{\bsnm{Zhang}, \binits{J.}},
\oauthor{\bsnm{Dong}, \binits{Z.}}, et al.:
A survey of large language models
(2023)
\end{botherref}
\endbibitem

\bibitem[\protect\citeauthoryear{Ali et~al.}{2023}]{Ali_Shamsan_Hezam_Mohammed_2023}
\begin{barticle}
\bauthor{\bsnm{Ali}, \binits{J.K.M.}},
\bauthor{\bsnm{Shamsan}, \binits{M.A.A.}},
\bauthor{\bsnm{Hezam}, \binits{T.A.}},
\bauthor{\bsnm{Mohammed}, \binits{A.A.}}:
\batitle{Impact of chatgpt on learning motivation: teachers and students' voices}.
\bjtitle{Journal of English Studies in Arabia Felix}
\bvolume{2}(\bissue{1}),
\bfpage{41}--\blpage{49}
(\byear{2023})
\end{barticle}
\endbibitem

\bibitem[\protect\citeauthoryear{Kasneci et~al.}{2023}]{kasneci2023chatgpt}
\begin{barticle}
\bauthor{\bsnm{Kasneci}, \binits{E.}},
\bauthor{\bsnm{Se{\ss}ler}, \binits{K.}},
\bauthor{\bsnm{K{\"u}chemann}, \binits{S.}},
\bauthor{\bsnm{Bannert}, \binits{M.}},
\bauthor{\bsnm{Dementieva}, \binits{D.}},
\bauthor{\bsnm{Fischer}, \binits{F.}},
\bauthor{\bsnm{Gasser}, \binits{U.}},
\bauthor{\bsnm{Groh}, \binits{G.}},
\bauthor{\bsnm{G{\"u}nnemann}, \binits{S.}},
\bauthor{\bsnm{H{\"u}llermeier}, \binits{E.}}, \betal:
\batitle{Chatgpt for good? on opportunities and challenges of large language models for education}.
\bjtitle{Learning and individual differences}
\bvolume{103},
\bfpage{102274}
(\byear{2023})
\end{barticle}
\endbibitem

\bibitem[\protect\citeauthoryear{Dong et~al.}{2024}]{dong2024}
\begin{bchapter}
\bauthor{\bsnm{Dong}, \binits{B.}},
\bauthor{\bsnm{Bai}, \binits{J.}},
\bauthor{\bsnm{Xu}, \binits{T.}},
\bauthor{\bsnm{Zhou}, \binits{Y.}}:
\bctitle{Large language models in education: A systematic review}.
In: \bbtitle{2024 6th International Conference on Computer Science and Technologies in Education (CSTE)},
pp. \bfpage{131}--\blpage{134}
(\byear{2024}).
\doiurl{10.1109/CSTE62025.2024.00031}
\end{bchapter}
\endbibitem

\bibitem[\protect\citeauthoryear{Ratnam et~al.}{2023}]{ratnam2023chatgpt}
\begin{botherref}
\oauthor{\bsnm{Ratnam}, \binits{M.}},
\oauthor{\bsnm{Sharma}, \binits{B.}},
\oauthor{\bsnm{Tomer}, \binits{A.}}:
Chatgpt: educational artificial intelligence.
International Journal, of advanced trends in computer science and engineering
\textbf{12}(2)
(2023)
\end{botherref}
\endbibitem

\bibitem[\protect\citeauthoryear{Memarian and Doleck}{2023}]{memarian2023chatgpt}
\begin{botherref}
\oauthor{\bsnm{Memarian}, \binits{B.}},
\oauthor{\bsnm{Doleck}, \binits{T.}}:
ChatGPT in education: Methods, potentials and limitations.
Elsevier
(2023)
\end{botherref}
\endbibitem

\bibitem[\protect\citeauthoryear{Farhi et~al.}{2023}]{farhi2023analyzing}
\begin{botherref}
\oauthor{\bsnm{Farhi}, \binits{F.}},
\oauthor{\bsnm{Jeljeli}, \binits{R.}},
\oauthor{\bsnm{Aburezeq}, \binits{I.}},
\oauthor{\bsnm{Dweikat}, \binits{F.F.}},
\oauthor{\bsnm{Al-shami}, \binits{S.A.}},
\oauthor{\bsnm{Slamene}, \binits{R.}}:
Analyzing the students' views, concerns, and perceived ethics about chat GPT usage.
Elsevier
(2023)
\end{botherref}
\endbibitem

\bibitem[\protect\citeauthoryear{Meyer et~al.}{2023}]{meyer2023chatgpt}
\begin{barticle}
\bauthor{\bsnm{Meyer}, \binits{J.G.}},
\bauthor{\bsnm{Urbanowicz}, \binits{R.J.}},
\bauthor{\bsnm{Martin}, \binits{P.C.}},
\bauthor{\bsnm{O’Connor}, \binits{K.}},
\bauthor{\bsnm{Li}, \binits{R.}},
\bauthor{\bsnm{Peng}, \binits{P.-C.}},
\bauthor{\bsnm{Bright}, \binits{T.J.}},
\bauthor{\bsnm{Tatonetti}, \binits{N.}},
\bauthor{\bsnm{Won}, \binits{K.J.}},
\bauthor{\bsnm{Gonzalez-Hernandez}, \binits{G.}}, \betal:
\batitle{Chatgpt and large language models in academia: opportunities and challenges}.
\bjtitle{BioData Mining}
\bvolume{16}(\bissue{1}),
\bfpage{20}
(\byear{2023})
\end{barticle}
\endbibitem

\bibitem[\protect\citeauthoryear{Tlili et~al.}{2023}]{tlili2023if}
\begin{barticle}
\bauthor{\bsnm{Tlili}, \binits{A.}},
\bauthor{\bsnm{Shehata}, \binits{B.}},
\bauthor{\bsnm{Adarkwah}, \binits{M.A.}},
\bauthor{\bsnm{Bozkurt}, \binits{A.}},
\bauthor{\bsnm{Hickey}, \binits{D.T.}},
\bauthor{\bsnm{Huang}, \binits{R.}},
\bauthor{\bsnm{Agyemang}, \binits{B.}}:
\batitle{What if the devil is my guardian angel: Chatgpt as a case study of using chatbots in education}.
\bjtitle{Smart learning environments}
\bvolume{10}(\bissue{1}),
\bfpage{15}
(\byear{2023})
\end{barticle}
\endbibitem

\bibitem[\protect\citeauthoryear{Bhullar et~al.}{2024}]{bhullar2024chatgpt}
\begin{botherref}
\oauthor{\bsnm{Bhullar}, \binits{P.S.}},
\oauthor{\bsnm{Joshi}, \binits{M.}},
\oauthor{\bsnm{Chugh}, \binits{R.}}:
ChatGPT in higher education-a synthesis of the literature and a future research agenda.
Springer
(2024)
\end{botherref}
\endbibitem

\bibitem[\protect\citeauthoryear{Albadarin et~al.}{2024}]{albadarin2024systematic}
\begin{barticle}
\bauthor{\bsnm{Albadarin}, \binits{Y.}},
\bauthor{\bsnm{Saqr}, \binits{M.}},
\bauthor{\bsnm{Pope}, \binits{N.}},
\bauthor{\bsnm{Tukiainen}, \binits{M.}}:
\batitle{A systematic literature review of empirical research on chatgpt in education}.
\bjtitle{Discover Education}
\bvolume{3}(\bissue{1}),
\bfpage{60}
(\byear{2024})
\end{barticle}
\endbibitem

\bibitem[\protect\citeauthoryear{Punar~Ozcelik and Yangin~Eksi}{2024}]{punar2024cultivating}
\begin{barticle}
\bauthor{\bsnm{Punar~Ozcelik}, \binits{N.}},
\bauthor{\bsnm{Yangin~Eksi}, \binits{G.}}:
\batitle{Cultivating writing skills: the role of chatgpt as a learning assistant—a case study}.
\bjtitle{Smart Learning Environments}
\bvolume{11}(\bissue{1}),
\bfpage{10}
(\byear{2024})
\end{barticle}
\endbibitem

\bibitem[\protect\citeauthoryear{Mogavi et~al.}{2024}]{mogavi2024chatgpt}
\begin{barticle}
\bauthor{\bsnm{Mogavi}, \binits{R.H.}},
\bauthor{\bsnm{Deng}, \binits{C.}},
\bauthor{\bsnm{Kim}, \binits{J.J.}},
\bauthor{\bsnm{Zhou}, \binits{P.}},
\bauthor{\bsnm{Kwon}, \binits{Y.D.}},
\bauthor{\bsnm{Metwally}, \binits{A.H.S.}},
\bauthor{\bsnm{Tlili}, \binits{A.}},
\bauthor{\bsnm{Bassanelli}, \binits{S.}},
\bauthor{\bsnm{Bucchiarone}, \binits{A.}},
\bauthor{\bsnm{Gujar}, \binits{S.}}, \betal:
\batitle{Chatgpt in education: A blessing or a curse? a qualitative study exploring early adopters’ utilization and perceptions}.
\bjtitle{Computers in Human Behavior: Artificial Humans}
\bvolume{2}(\bissue{1}),
\bfpage{100027}
(\byear{2024})
\end{barticle}
\endbibitem

\bibitem[\protect\citeauthoryear{Stojanov et~al.}{2024}]{stojanov2024university}
\begin{barticle}
\bauthor{\bsnm{Stojanov}, \binits{A.}},
\bauthor{\bsnm{Liu}, \binits{Q.}},
\bauthor{\bsnm{Koh}, \binits{J.H.L.}}:
\batitle{University students’ self-reported reliance on chatgpt for learning: A latent profile analysis}.
\bjtitle{Computers and Education: Artificial Intelligence}
\bvolume{6},
\bfpage{100243}
(\byear{2024})
\end{barticle}
\endbibitem

\bibitem[\protect\citeauthoryear{Al~Murshidi et~al.}{2024}]{al2024understanding}
\begin{barticle}
\bauthor{\bsnm{Al~Murshidi}, \binits{G.}},
\bauthor{\bsnm{Shulgina}, \binits{G.}},
\bauthor{\bsnm{Kapuza}, \binits{A.}},
\bauthor{\bsnm{Costley}, \binits{J.}}:
\batitle{How understanding the limitations and risks of using chatgpt can contribute to willingness to use}.
\bjtitle{Smart Learning Environments}
\bvolume{11}(\bissue{1}),
\bfpage{36}
(\byear{2024})
\end{barticle}
\endbibitem

\bibitem[\protect\citeauthoryear{Qu et~al.}{2024}]{qu2024disciplinary}
\begin{barticle}
\bauthor{\bsnm{Qu}, \binits{Y.}},
\bauthor{\bsnm{Tan}, \binits{M.X.Y.}},
\bauthor{\bsnm{Wang}, \binits{J.}}:
\batitle{Disciplinary differences in undergraduate students' engagement with generative artificial intelligence}.
\bjtitle{Smart Learning Environments}
\bvolume{11}(\bissue{1}),
\bfpage{51}
(\byear{2024})
\end{barticle}
\endbibitem

\bibitem[\protect\citeauthoryear{Velasquez-Astuhuaman and Libaque-Saenz}{2024}]{velasquez2024main}
\begin{barticle}
\bauthor{\bsnm{Velasquez-Astuhuaman}, \binits{P.R.}},
\bauthor{\bsnm{Libaque-Saenz}, \binits{C.F.}}:
\batitle{Main factors driving the use of chatgpt by undergraduate students: An exploratory and preliminary study in peru}.
\bjtitle{Issues in Information Systems}
\bvolume{25}(\bissue{1}),
\bfpage{402}--\blpage{418}
(\byear{2024})
\end{barticle}
\endbibitem

\bibitem[\protect\citeauthoryear{Tiwari et~al.}{2024}]{tiwari2024drives}
\begin{barticle}
\bauthor{\bsnm{Tiwari}, \binits{C.K.}},
\bauthor{\bsnm{Bhat}, \binits{M.A.}},
\bauthor{\bsnm{Khan}, \binits{S.T.}},
\bauthor{\bsnm{Subramaniam}, \binits{R.}},
\bauthor{\bsnm{Khan}, \binits{M.A.I.}}:
\batitle{What drives students toward chatgpt? an investigation of the factors influencing adoption and usage of chatgpt}.
\bjtitle{Interactive Technology and Smart Education}
\bvolume{21}(\bissue{3}),
\bfpage{333}--\blpage{355}
(\byear{2024})
\end{barticle}
\endbibitem

\bibitem[\protect\citeauthoryear{Foroughi et~al.}{2024}]{foroughi2024determinants}
\begin{barticle}
\bauthor{\bsnm{Foroughi}, \binits{B.}},
\bauthor{\bsnm{Senali}, \binits{M.G.}},
\bauthor{\bsnm{Iranmanesh}, \binits{M.}},
\bauthor{\bsnm{Khanfar}, \binits{A.}},
\bauthor{\bsnm{Ghobakhloo}, \binits{M.}},
\bauthor{\bsnm{Annamalai}, \binits{N.}},
\bauthor{\bsnm{Naghmeh-Abbaspour}, \binits{B.}}:
\batitle{Determinants of intention to use chatgpt for educational purposes: Findings from pls-sem and fsqca}.
\bjtitle{International Journal of Human--Computer Interaction}
\bvolume{40}(\bissue{17}),
\bfpage{4501}--\blpage{4520}
(\byear{2024})
\end{barticle}
\endbibitem

\bibitem[\protect\citeauthoryear{Strzelecki}{2023}]{strzelecki2023use}
\begin{botherref}
\oauthor{\bsnm{Strzelecki}, \binits{A.}}:
To use or not to use ChatGPT in higher education? A study of students’ acceptance and use of technology.
Taylor \& Francis
(2023)
\end{botherref}
\endbibitem

\bibitem[\protect\citeauthoryear{Wang and Degol}{2017}]{wang2017gender}
\begin{barticle}
\bauthor{\bsnm{Wang}, \binits{M.-T.}},
\bauthor{\bsnm{Degol}, \binits{J.L.}}:
\batitle{Gender gap in science, technology, engineering, and mathematics (stem): Current knowledge, implications for practice, policy, and future directions}.
\bjtitle{Educational psychology review}
\bvolume{29},
\bfpage{119}--\blpage{140}
(\byear{2017})
\end{barticle}
\endbibitem

\bibitem[\protect\citeauthoryear{Naka and Naoi}{1995}]{naka1995effect}
\begin{barticle}
\bauthor{\bsnm{Naka}, \binits{M.}},
\bauthor{\bsnm{Naoi}, \binits{H.}}:
\batitle{The effect of repeated writing on memory}.
\bjtitle{Memory \& cognition}
\bvolume{23},
\bfpage{201}--\blpage{212}
(\byear{1995})
\end{barticle}
\endbibitem

\bibitem[\protect\citeauthoryear{Askvik et~al.}{2020}]{van2020importance}
\begin{barticle}
\bauthor{\bsnm{Askvik}, \binits{E.O.}},
\bauthor{\bsnm{Weel}, \binits{F.R.R.}},
\bauthor{\bsnm{Meer}, \binits{A.L.H.}}:
\batitle{The importance of cursive handwriting over typewriting for learning in the classroom: A high-density eeg study of 12-year-old children and young adults.}
\bjtitle{Frontiers in Psychology}
\bvolume{11},
\bfpage{1810}--\blpage{1810}
(\byear{2020})
\end{barticle}
\endbibitem

\bibitem[\protect\citeauthoryear{Biswas}{2023}]{biswas2023potential}
\begin{barticle}
\bauthor{\bsnm{Biswas}, \binits{S.S.}}:
\batitle{Potential use of chat gpt in global warming}.
\bjtitle{Annals of biomedical engineering}
\bvolume{51}(\bissue{6}),
\bfpage{1126}--\blpage{1127}
(\byear{2023})
\end{barticle}
\endbibitem

\bibitem[\protect\citeauthoryear{George et~al.}{2023}]{george2023environmental}
\begin{barticle}
\bauthor{\bsnm{George}, \binits{A.S.}},
\bauthor{\bsnm{George}, \binits{A.H.}},
\bauthor{\bsnm{Martin}, \binits{A.G.}}:
\batitle{The environmental impact of ai: a case study of water consumption by chat gpt}.
\bjtitle{Partners Universal International Innovation Journal}
\bvolume{1}(\bissue{2}),
\bfpage{97}--\blpage{104}
(\byear{2023})
\end{barticle}
\endbibitem

\bibitem[\protect\citeauthoryear{Zhu et~al.}{2023}]{zhu2023chatgpt}
\begin{barticle}
\bauthor{\bsnm{Zhu}, \binits{J.-J.}},
\bauthor{\bsnm{Jiang}, \binits{J.}},
\bauthor{\bsnm{Yang}, \binits{M.}},
\bauthor{\bsnm{Ren}, \binits{Z.J.}}:
\batitle{Chatgpt and environmental research}.
\bjtitle{Environmental Science \& Technology}
\bvolume{57}(\bissue{46}),
\bfpage{17667}--\blpage{17670}
(\byear{2023})
\end{barticle}
\endbibitem

\end{thebibliography}

\end{document}